\documentclass[%
  aps,
  prd,
  twocolumn,
  superscriptaddress,
  floatfix,
  nobibnotes,
  nofootinbib,
]
{revtex4-2}

\usepackage[utf8]{inputenc}
\usepackage[german,english]{babel}

\usepackage{xargs}

\usepackage{graphicx}
\usepackage{color}
\usepackage{cancel}
\usepackage{lineno}
\usepackage{xspace}
\usepackage{siunitx}
\usepackage{dcolumn}
\usepackage{bm}
\usepackage[normalem]{ulem}
\usepackage{url}
\usepackage{float}
\usepackage{orcidlink}

\usepackage[fixamsmath,disallowspaces]{mathtools}
\usepackage{amssymb}
\usepackage[all,warning]{onlyamsmath}
\usepackage{aligned-overset}
\usepackage{bbm}
\usepackage{slashed}
\usepackage{braket}
\usepackage{tensor}

\usepackage{rotating}

\usepackage{nicefrac}

\usepackage{svn-multi}
\usepackage{enumitem}
\usepackage{wrapfig}
\usepackage{xcolor}
\usepackage{booktabs}
\usepackage{multirow}

\usepackage{hyperref}
\hypersetup{
  colorlinks=true,
  citecolor=blue,
  citebordercolor=red,
  linktoc=all,
  linkcolor=blue,
  urlcolor=blue
}

\usepackage[capitalise]{cleveref}
\AddToHook{cmd/appendix/before}{
  
  \crefalias{section}{appendix}
  \crefalias{subsection}{appendix}
}

\crefformat{equation}{#2Eq.~(#1)#3}
\crefformat{appendix}{#2App.~#1#3}
\crefformat{section}{#2Sec.~#1#3}
\crefformat{table}{#2Tab.~#1#3}
\crefformat{figure}{#2Fig.~#1#3}

\crefmultiformat{section}{#2Secs.~#1#3}{ and~#2#1#3}{, #2#1#3}{
and~#2#1#3}
\crefmultiformat{equation}{#2Eqs.~(#1)#3}{ and~#2(#1)#3}{, #2(#1)#3}{
and~#2(#1)#3}
\crefmultiformat{figure}{#2Figs.~#1#3}{ and~#2#1#3}{, #2#1#3}{
and~#2#1#3}

\crefrangeformat{equation}{#3Eqs.~(#1)#4 to #5(#2)#6}
\crefrangeformat{appendix}{#3Apps.~#1#4 to #5#2#6}

\Crefformat{figure}{#2Fig.~#1#3}

\usepackage[acronym]{glossaries-extra}
\glssetcategoryattribute{acronym}{nohyperfirst}{true}
\setabbreviationstyle[acronym]{long-short}

\newacronym{QCD}{QCD}{quantum chromodynamics}
\newacronym{RG}{RG}{renormalization group}
\newacronym{FRG}{FRG}{functional renormalization group}
\newacronym{MFA}{MFA}{mean-field approximation}
\newacronym{RPA}{RPA}{random-phase approximation}
\newacronym{RGC}{RGC}{RG consistency}
\newacronym{LPA}{LPA}{local-potential approximation}
\newacronym{mLPA}{mLPA}{\textit{mesonic} local-potential approximation}
\newacronym{DSE}{DSE}{Dyson-Schwinger equations}
\newacronym{UV}{UV}{ultraviolet}
\newacronym{IR}{IR}{infrared}
\newacronym{EoS}{EoS}{equation of state}
\newacronym{QMD}{QMD}{quark--meson--diquark}
\newacronym{NJL}{NJL}{Nambu--Jona-Lasinio}
\newacronym{2SC}{2SC}{two-flavor color-superconducting}
\newacronym{CFL}{CFL}{color-flavor-locked}
\newacronym{BCS}{BCS}{Bardeen--Cooper--Schrieffer}

\newcommand{\ccite}[1]{Ref.~\cite{#1}}
\newcommand{\ccites}[1]{Refs.~\cite{#1}}

\usepackage{natbib}
\newcommand{\twofigs}{0.49\textwidth}

\newcommand{\Tr}{\mathrm{Tr}}
\newcommand{\tr}{\mathrm{tr}}

\newcommand{\sqvec}[1]{\ensuremath{ \vec{#1}^{\, 2}}}

\DeclareMathOperator{\diag}{diag}

\newcommand*{\tp}{\mathsf{T}}

\newcommand{\gorkov}{Gor'kov\xspace}

\newcommand{\da}{\mathfrak{a}}
\newcommand{\db}{\mathfrak{b}}
\newcommand{\dc}{\mathfrak{c}}
\newcommand{\dd}{\mathfrak{d}}
\newcommand{\de}{\mathfrak{e}}
\newcommand{\df}{\mathfrak{f}}

\newcommand{\MeV}{\,\mathrm{MeV}}
\newcommand{\GeV}{\,\mathrm{GeV}}

\newcommand{\JLU}{%
  Institut f\"{u}r Theoretische Physik, %
  Justus-Liebig-Universit\"{a}t Gie\ss{}en, %
  35392 Gie\ss{}en, %
  Germany%
}

\newcommand{\HFHF}{%
  Helmholtz Forschungsakademie Hessen f\"{u}r FAIR (HFHF), %
  GSI Helmholtzzentrum f\"{u}r Schwerionenforschung, %
  Campus Gie\ss{}en, %
  35392 Gie\ss{}en, %
  Germany%
}

\begin{document}

\title{Diquark Correlators and Phase Structure in the Quark-Meson-Diquark Model
beyond Mean Field}

\author{Ugo Mire \orcidlink{0009-0009-2345-691X}}
\email{ugo.louis.tryphon.mire@physik.uni-giessen.de}
\affiliation{\JLU}

\author{Bernd-Jochen Schaefer \orcidlink{0000-0003-0659-2679}}
\email{bernd-jochen.schaefer@uni-giessen.de}
\affiliation{\JLU}
\affiliation{\HFHF}

\begin{abstract}
  A comprehensive study of the phase structure of the two-flavor
  quark–meson–diquark model is presented within the nonperturbative
  functional renormalization group framework. The influence of
  mesonic fluctuations beyond the mean-field approximation is
  investigated, and two-point functions of the diquark fields are
  computed at finite real-time frequencies. Renormalization group
  consistency of the effective potential is ensured in order to
  avoid cutoff artifacts.  Substantial modifications of the phase
  structure are found once mesonic fluctuations are included, and
  for sufficiently strong diquark couplings the dynamics become
  dominated by diquark condensation. These effects are elucidated
  through an analysis of the diquark pole mass and the Silver-Blaze
  property.
\end{abstract}

\maketitle

\section{Introduction}
\label{sec:introduction}

The microscopic theory governing strong-interacting matter is
\gls{QCD}, whose asymptotically free running coupling
increases logarithmically with decreasing momentum scale. This scale
dependence implies that at low energies the strong coupling becomes
large, leading to a rich spectrum of bound states and the dynamical
emergence of key non-perturbative phenomena such as spontaneous chiral
symmetry breaking and confinement.

While at sufficiently high energy or baryon densities \gls{QCD} becomes
amenable to perturbative methods for weakly-interacting quarks and
gluons due to asymptotic freedom, low-energy \gls{QCD} requires
nonperturbative techniques to account for the dynamical changes of the
relevant degrees of freedom.

In the low-density and low-temperature region, hadronic effective field theories
incorporating chiral symmetry and confinement provide a successful description
of equilibrium thermodynamics and the hadronic spectrum, see the review
\cite{Hebeler:2020ocj}. As baryon density increases, several qualitative changes
in the ground state of \gls{QCD} are expected to occur: chiral symmetry is
anti\-cipated to be restored, quark matter may exhibit various
color-superconducting phases, and more exotic possibilities such as crystalline
or inhomogeneous phases have been proposed \cite{Alford:1997zt, Alford:2001dt,
Alford:2007xm, Fukushima:2010bq, Anglani:2013gfu, Buballa:2014tba, Fu:2019hdw,
Pisarski:2021qof, Fu:2024rto, Motta:2024rvk, Schmitt:2025cqi,
Pawlowski:2025jpg}. Mapping out these regimes and determining the corresponding
\gls{EoS} remains a major challenge, largely because the sign problem prevents
lattice \gls{QCD} from providing first-principles results at finite chemical
potential, see, e.g., \ccites{HotQCD:2018pds, Borsanyi:2020fev, Dini:2021hug,
Namekawa:2022liz}.

At finite baryon density, functional nonperturbative methods such as
the \gls{FRG} and \gls{DSE} constitute the only first-principles
approaches capable of directly accessing \gls{QCD} thermodynamics.
Nevertheless, lattice results on the phase
structure and on conserved-charge fluctuations remain indispensable
benchmarks for functional \gls{QCD} methods at
$\mu_B =0$, see the recent review \cite{Aarts:2023vsf}.
Functional methods have now reached quantitative control up to the
regime $\mu_B/T \lesssim 4.5$, enabling studies of the \gls{QCD}
\gls{EoS} and phase diagram at moderate densities in a systematic
manner; see, e.g.,~\ccites{Fischer:2018sdj,Fu:2022gou,
  Dupuis:2020fhh, Fischer:2026uni, Pawlowski:2025jpg}.

The \gls{FRG} framework employed in this work builds upon these advances by
using a systematically improvable low-energy truncation that
incorporates key insights from first-principles \gls{QCD} at finite
temperature and density \cite{Braun:2007bx, Braun:2014ata, Mitter:2014wpa, Cyrol:2016tym,
Cyrol:2017ewj, Braun:2021uua}.  A central result of
these studies is that the gluonic sector of \gls{QCD} decouples efficiently
at momenta around 1 GeV, owing to the dynamical gluon mass gap. This
decoupling justifies the construction of quantitatively reliable
low-energy effective descriptions in which the dominant infrared
physics is encoded in quark, meson, and diquark degrees of freedom
\cite{Fu:2019hdw}.  At even lower energies, most hadronic modes
decouple as well, further supporting the success of chiral effective
field theories, such as quark–meson models, in describing
low-temperature and low-density \gls{QCD}.

Consequently, low-energy \gls{QCD} can be accurately captured without
explicitly resolving the full gluonic dynamics, provided that the
matching to \gls{QCD} in the ultraviolet is performed consistently and
that renormalization-group constraints are
respected. This motivates the use of \gls{QMD}–type
  models \cite{Braun:2018svj,Andersen:2024qus,Gholami:2025afm,
    Andersen:2025ezj, Andersen:2026xrf}, which have also been
  introduced in the context of two-color \gls{QCD}
  \cite{Kogut:1999iv,Andersen:2010vu,Strodthoff:2011tz,Khan:2015puu}.
  Supplemented by \gls{FRG} evolution, these models provide controlled
  low-energy frameworks for investigating thermodynamic properties and
  the equation of state at finite temperature and density. Such
models allow one to incorporate chiral symmetry breaking, diquark
pairing, and potential color-superconducting phases while maintaining
a close connection to underlying \gls{QCD} dynamics through their
\gls{RG} structure.

The \gls{FRG} provides a particularly suitable framework to
investigate this decoupling and the emergence of hadronic bound states
from \gls{QCD}, see \ccites{Evans:1998nf, Schafer:1998na} for early
works.  Within this approach, composite fields can be incorporated
dynamically via dynamical hadronization \cite{Gies:2001nw},
yielding low-energy effective theories that remain consistently
embedded in the underlying quark-gluon dynamics. In practice, this
procedure leads to Polyakov-loop-extended \gls{NJL}- or
quark-meson-type models, with additional diquark channels emerging
once quantum, thermal and density fluctuations are taken into account
\cite{Fu:2021oaw,Fukushima:2021ctq,Braun:2021uua}. Recently, related \gls{FRG} studies have explored
quark-diquark models as effective descriptions of baryonic degrees of
freedom, including an extrapolation to a \gls{QMD} model, thereby
providing a complementary description of bound-state formation in the
low-energy regime of \gls{QCD} \cite{Stoll:2025jor, Gholami:2025afm}.

Motivated by these developments, we employ here an effective low-energy
truncation that explicitly includes diquark degrees of freedom and is
tailored to intermediate and high quark densities. This construction
is guided by recent \gls{FRG} results on \gls{QCD} quark-gluon dynamics
\cite{Leonhardt:2019fua, Braun:2022olp}, thereby providing a first
step towards a \gls{QCD}-constrained \gls{FRG} study of the \gls{QMD}
model at finite temperature and density.

Specifically, we investigate a two-flavor \gls{QMD} model truncation
as a low-energy realisation of the \gls{2SC} phase of \gls{QCD} at
moderate baryon density. At asymptotically large densities, the
\gls{QCD} ground state is expected to be in the \gls{CFL} phase
\cite{Alford:2001dt,Alford:2007xm,Schmitt:2025cqi}.
At lower chemical potentials, differences in the effective quark
masses become more pronounced, leading to mismatched Fermi momenta and
disfavouring \gls{CFL} pairing. This, in turn, modifies the conditions for
Cooper pairing among quarks and can change the preferred pairing
patterns. In the density regime, where strange quarks effectively
decouple, phases such as the \gls{2SC} phase, may therefore become
favored \cite{Alford:2001dt}.

Recent \gls{FRG} analyses revealed that at low densities the dynamics
is mostly governed by the (pseudo)scalar mesonic channel associated
with chiral symmetry breaking, while at higher densities attractive
diquark correlations become more dominant, cf~\ccites{Braun:2011pp,
  Braun:2019aow}, in agreement with weak-coupling results at
asymptotically large densities \cite{Son:1998uk,Pisarski:1999bf}.  We
present a detailed study of the \gls{QMD} model at finite temperature
and density in order to clarify how thermal and density fluctuations
influence the competition between chiral and diquark ordering and
consequently, the resulting phase structure.

This paper is organized as follows. In \cref{sec:low-energy-real} we briefly
introduce the \gls{QMD} model, with emphasis on some of its symmetry properties.
In \cref{sec:funct-renorm-group} we present our \gls{FRG} approach for deriving
the thermodynamic grand potential, discuss its relation to the often employed
\gls{MFA}, and explain how cutoff artifacts are treated via \gls{RGC}. In
\cref{sec:corr-funct-at} we provide details on the computation of the
two-point functions of the diquark field in vacuum.
Numerical results are summarized in \cref{sec:numerical-results}. We conclude in
\cref{sec:summary} with a summary and outlook. Several technical details are
collected in \crefrange{app:regulator-functions}{app:uv-param-numer}.

\section{Low-energy Realization of QCD}
\label{sec:low-energy-real}

The two-flavor \gls{QMD} model is an effective low-energy description
of \gls{QCD} that captures the essential aspects of chiral symmetry
breaking and color superconductivity. We begin with a brief analysis
of its symmetries, emphasizing the transformation properties of the
diquark field under color and flavor rotations.

\subsection{The quark-meson-diquark model}
\label{sec:quark-meson-diquark}

The Euclidean classical action of the $N_f= 2$ flavor and $N_c=3$
color \gls{QMD} model at finite temperature
$\beta = 1/T$ and quark chemical potential $\mu$ reads
\begin{widetext}
  \begin{equation} \label{eq:classical_action}
    \begin{aligned}
      S[\bar{q},q,\phi,  {\Delta}, {\Delta}^{*}] = &
      \int_0^\beta\!\! dx_0 \int\!\! d^3x \bigg\{
        \bar{q}\left[
          \slashed{\partial} - \mu \gamma_4
          + g_\phi \left(
            \sigma + i \gamma_5 \vec{\pi} \cdot \vec{\tau}
          \right)
        \right] q
        + \frac{1}{2} g_\Delta \left(
          \Delta_\da \bar{q} \gamma_5 \tau_2 i\epsilon_{\da} C \bar{q}^\tp
          - \Delta_\da^* q^\tp C \gamma_5 \tau_2 i\epsilon_{\da} q
        \right)                           \\[1ex] &
        + \frac{1}{2}(\partial_\mu \phi) (\partial_\mu \phi)
        + (\partial_\mu + 2\mu\delta_{\mu 4})\Delta_\da^*
        (\partial_\mu - 2\mu\delta_{\mu 4})\Delta_\da
        + U(\phi^2, |\Delta|^2)- h\sigma
      \bigg\} \; .
    \end{aligned}
  \end{equation}
\end{widetext}
It describes the interactions between quark fields, $q$ and $\bar{q}$,
and effective degrees of freedom associated with the $\sigma$- and
$\vec{\pi}$-mesons, as well as diquark fields $\Delta$ and
$\Delta^{*}$.

The mesonic fields are grouped into the $O(4)$-symmetric chiral field
$\phi = (\sigma, \vec{\pi})$, which provides a convenient
basis for the chirally invariant quantity
$\phi^2 = \sigma^2 + \vec{\pi}^2$.  The potential
$U(\phi^2, |\Delta|^2)$ encodes the meson-meson, diquark-meson as well
as diquark-diquark interactions consistent with the underlying chiral
and color symmetries.  The diquark invariant is given by
$|\Delta|^2 = \sum_\da \Delta_\da^{*} \Delta_\da$, where $\da$ runs
over the $N_c=3$ antisymmetric color generators in the fundamental
representation.

In the present  work, we restrict ourselves to the minimal truncation
\begin{equation}
  \label{eq:2}  
  U(\phi^2, |\Delta|^2) = \frac{1}{2} m^2_\phi \phi^2 
  + \frac{1}{4} \lambda_\phi \phi^4 
  + m^2_\Delta |\Delta|^2 \; ,
\end{equation}
which comprises quadratic mass terms for the mesonic and diquark
fields as well as a quartic self-interaction in the mesonic sector.
The quark fields $q$ and $\bar{q}$ are Dirac spinors carrying color
and flavor indices and transform in the fundamental representation of
the color group $SU(3)_c$. The two Yukawa couplings $g_{\phi}$ and
$g_{\Delta}$ describe the flavor- and color-blind quark-meson and
quark-diquark couplings. The explicit chiral symmetry breaking is
implemented with a linear term in the radial sigma direction with
strength $h$, generating finite current quark masses and,
consequently, non-vanishing pion masses. As a result, the chiral phase
transition at vanishing density becomes a smooth crossover rather than
a sharp second-order transition.

The charge-conjugated quark spinors $q^C = C \bar{q}^\tp$ and
$\bar{q}^C = q^\tp C$ are defined using the charge-conjugation matrix
$C=\gamma_2 \gamma_4$ in Euclidean space, where the Dirac matrices
$\gamma_{\mu}$ are Hermitian, $\gamma_\mu^\dagger = \gamma_\mu$, and
satisfy the normalization
$\{\gamma_\mu,\gamma_\nu\} = 2\delta_{\mu\nu}$. The Pauli matrices
$\vec{\tau} = (\tau_1, \tau_2, \tau_3)$ act in flavor space, while the
totally antisymmetric tensor $(\epsilon_{\da})_{\db\dc} = \epsilon_{\da\db\dc}$
acts in color space.

The composite, complex-valued scalar diquark fields $\Delta_\da$ and
$\Delta^*_\da$ are treated as components of a vector in color space.
These fields carry the quantum numbers of scalar diquarks with
$J^P = 0^+$ and are represented as
\begin{equation}
  \label{eq:diquark_pairing_pattern}
  \Delta_\da \sim q^\tp C \tau_2 i \epsilon_{\da} \gamma_5 q \; .
\end{equation}
In the following, Fraktur indices denote color, while flavor indices are
suppressed for clarity.

Owing to the nonvanishing baryon charge of the
  diquark field, its kinetic term exhibits a qualitatively different
  structure from that of the mesonic sector. The (composite) diquarks
  carry baryon number $B=2/3$, and the quark chemical potential
  couples to baryon number as the temporal component of an external
  $U(1)_B$ gauge field. Accordingly, the ordinary derivative is
  promoted to a covariant one
\begin{equation}
\label{eq:71}
\partial_{\mu} \to \partial_{\mu} \mp 2\mu \delta_{\mu4}\ ,
\end{equation}
which modifies only the temporal component and incorporates the
finite-density background. The opposite signs for $\Delta_\da$ and
$\Delta^{*}_\da$ ensure invariance under global $U(1)_B$
transformations and implement the coupling to the conserved charge. In
contrast, the mesonic fields carry zero baryon number and therefore do
not couple directly to the chemical potential at the kinetic
level. The resulting diquark kinetic term thus provides the minimal
realization of finite-density effects for charged composite bosons,
leading to a $\mu$-dependent dispersion relation.

In the present truncation, we neglect wave-function renormalization
effects for all fields, in contrast to \ccite{Gholami:2025afm}, where
the diquark wave-function renormalization was argued to play an
important role in controlling medium-induced divergences.
Here, however, cutoff artifacts are treated by means of RG-consistent
flows, as we will discuss in \cref{sec:renorm-group-cons}. As
demonstrated in \ccite{Gholami:2025afm}, this procedure can be used
instead of a standard renormalization scheme, such that the
counterterms typically associated with diquark wave-function
renormalization are generated dynamically within the flow.

\subsection{Symmetries of the model}
\label{sec:symmetries-model}

For two quark flavors and in the chiral limit, the \gls{QMD} model
exhibits the global symmetry
$SU(3)_c \times SU(2)_L \times SU(2)_R\times U(1)_B$, corresponding to
color, chiral flavor, and baryon number conservation\footnote{The axial $U(1)_A$
    symmetry is assumed to be maximally broken by the QCD axial
    anomaly, as expected from instanton-induced interactions.}. The
meson fields form a chiral multiplet under $SU(2)_L \times SU(2)_R$,
while the scalar diquark fields transform as color antitriplets and
carry nonzero baryon charge. In the vacuum, a finite expectation value
of the sigma field spontaneously breaks chiral symmetry according to
$SU(2)_L \times SU(2)_R \to SU(2)_V$, generating constituent quark
masses and three (pseudo-)Nambu-Goldstone pions.

At sufficiently large baryon chemical potential, the formation of a diquark
condensate becomes energetically favored, leading to spontaneous breaking of
baryon number symmetry $U(1)_B$. Simultaneously, the global color symmetry is
broken according to $SU(3)_c \to SU(2)_c$, corresponding to the \gls{2SC} phase.
Due to the global color symmetry breaking, the spontaneous breaking produces
($8-3$) additional Goldstone modes, leaving one massive radial diquark mode.  In
a local gauge theory these modes would instead correspond to ($8-3$) massive
gauge bosons via the Anderson-Higgs mechanism.  Thus, depending on the model
parameters, chiral and diquark condensates may coexist over a range of
densities, resulting in simultaneous breaking of chiral, baryon, and color
symmetries. The \gls{QMD} model captures the transition from a chirally broken
hadronic phase to a color-superconducting phase driven by diquark condensation.

The chiral (pseudo)scalar interaction sector is a color singlet and
therefore preserves the global color symmetry of the
model. Conversely, the diquark interaction channel is invariant under
chiral $SU(2)_L \times SU(2)_R$ transformations and does not modify
the chiral symmetry structure. This separation becomes manifest upon
inspecting the transformation properties of the corresponding
interaction terms.

The diquark fields and their conjugates are invariant under chiral
transformations of the quark fields,
\begin{equation}
  \label{eq:24}
  q \to U_{\chi} q\; ,
  \quad
  U_{\chi} \in SU(2)_L \otimes SU(2)_R
\end{equation}
and thus transform as chiral singlets.
To see this, introduce the chiral projectors
\begin{equation}
  \label{eq:31}
  P_{L/ R}= (1 \mp \gamma_5) / 2\; , \quad    U_\chi = U_L P_L + U_R P_R \; ,
\end{equation}
with
\begin{equation}
  \label{eq:33}
  q_{L / R} \to U_{L / R} \, q_{L / R}\; ,
  \quad
  U_{L / R} = \exp\left(i\vec\theta_{L/ R} \cdot \vec\tau\right) \; ,
\end{equation}
then the diquark fields transform as
\begin{equation}
  \label{eq:34}
  \begin{aligned}
    \Delta_\da \to & \ q^\tp (U_L^\tp P_L + U_R^\tp P_R) C
    \tau_2 i \epsilon_{\da} \gamma_5 (U_L P_L + U_R P_R) q    \\[1ex]
    \to          & \ q^\tp
    C
    (U_L^\tp \tau_2 U_L P_L + U_R^\tp \tau_2 U_R P_R)
    i \epsilon_{\da} \gamma_5
    q \\[1ex]
    \to          & \ q^\tp C \tau_2 i \epsilon_{\da} \gamma_5 q
    = \Delta_\da \, .
  \end{aligned}
\end{equation}
Here the relation $\vec{\tau}=-\tau_2 \vec{\tau}^\tp \tau_2$, a direct
consequence of the pseudo-reality of the Pauli matrices
$\vec{\tau}^{*}= -\tau_2 \vec{\tau} \tau_2$, has been used.  Hence, a
finite diquark condensate does not break chiral symmetry.

In contrast, the diquark fields transform non-trivially under color
transformations as a color anti-triplet.  For a color transformation
of the quark fields,
\begin{equation}
  \label{eq:quark_color_transformation}
  q \to U q\; ,
  \quad
  U \in SU(3)_c \; ,
\end{equation}
the diquark fields transform as
\begin{equation}
  \label{eq:26}
  \Delta_\da \to q^\tp C \tau_2 i \left( U^\tp \epsilon_{\da}
  U \right) \gamma_5 q \; .
\end{equation}
Using for any $M \in SU(3)$ the identity
\begin{equation}
  \label{eq:43}
  \epsilon_{\da\db\dc} M_{\da\dd} M_{\db\de} M_{\dc\df} = \epsilon_{\dd\de\df}\; ,
\end{equation}
one finds
\begin{equation}
  \label{eq:63}
  U^\tp \epsilon_{\da} U = U_{\da\db}^* \epsilon_{\db}\ .
\end{equation}
Hence, the diquark fields transform according to the color
anti-triplet representation,
\begin{equation}
  \label{eq:27}
  \Delta_\da \to U_{\da\db}^* \Delta_\db \ , \quad \da=1,2,3\ .
\end{equation}

Finally, let us comment on the baryon number symmetry.  Under a global
$U(1)_B$ rotation with angle $\theta$, the diquark field transforms as
\begin{equation}
  \Delta_\da \to e^{2/3 i \theta} \Delta_\da \; ,
\end{equation}
with baryon charge $B=2/3$. A finite diquark condensate explicitly
breaks the naive baryon number symmetry. However, the condensate locks
color and baryon transformations such that a linear combination of the
original $U(1)_B$ generator and a generator of the color group remains
unbroken. This residual symmetry can be interpreted as a rotated (or
shifted) baryon number \cite{Buballa:2003qv}, implying that no
independent global symmetry is actually broken and thus no additional
Goldstone mode arises.

\section{Functional Renormalization Group}
\label{sec:funct-renorm-group}

The \gls{FRG} is a powerful method which allows one to incorporate both bosonic
and fermionic fluctuations in a nonperturbative manner. In the past, the
Euclidean formulation of the \gls{FRG} has been applied successfully to
investigate the phase structure of quark-meson models and variants thereof
\cite{Berges:1997eu, Schaefer:2004en, Schaefer:2006ds, Strodthoff:2011tz,
Khan:2015puu, Rennecke:2016tkm, Resch:2017vjs, Otto:2020hoz, Grossi:2021ksl,
Ihssen:2023xlp}.

For our analysis at moderate and high densities including low
temperatures we employ a standard 1PI flow pioneered by Wetterich
\cite{Wetterich:1992yh}. It is a functional differential equation for
the quantum average effective action $\Gamma_k$ which depends on the
infrared \gls{RG} cutoff scale $k$
\begin{equation}
  \label{eq:WetterichEq}
  \partial_t \Gamma_k [\Phi]= \frac{1}{2}\text{STr}\left[
    \Big( \Gamma^{(2)}_k [\Phi]+ R_k
    \Big)^{-1} \partial_t R_k
  \right] \; ,
\end{equation}
and has a simple one-loop structure in terms of the full field- and
momentum-dependent propagator. The infrared cutoff is introduced
through the regulator function $R_k$, which represents an \gls{IR} mass term
that vanishes in the \gls{IR} as $k\to 0$, see
\cref{app:regulator-functions} for details. Quantum fluctuations below
the cutoff scale $p^2 \lesssim k^2$ are suppressed and are integrated
out by lowering the scale $k$. The logarithmic scale derivative
$t=\ln k/\Lambda$ implements the so-called \gls{RG}-time with the reference
scale to be the \gls{UV} cutoff scale of the flow $\Lambda$ and runs from 0
to $-\infty$.  The physical theory is obtained in the infrared limit
$k\to 0$ with the physical effective action
$\Gamma[\Phi] \equiv \Gamma_{k=0}[\Phi]$.

The superfield $\Phi$ collects all fundamental fields of the theory,
and accordingly, the regulator function $R_k$ is understood as a
matrix in this field space.  The supertrace $\text{STr}$ comprises
momentum integrations and sums over all internal indices, such as
Lorentz indices and field species, including the appropriate minus
signs for fermionic contributions.  The $n$-point field derivatives of
the flow of the average effective action are defined as
\begin{equation}
  \label{eq:7}
  \Gamma^{(n)}_{\Phi_1 \ldots \Phi_n, k} (p_1, \ldots, p_{n-1})
  =\int_{p_n}\frac{ \delta^n \Gamma_k
  [\Phi]} {\delta  \Phi_1 (p_1) \cdots \delta \Phi_n (p_n)} \; ,
\end{equation}
which leads to an infinite coupled tower of flows of its $n$-th
moments. The momentum integration
$\int_{p_n} \equiv \int \frac{d^4 p_n}{(2\pi)^4}$ eliminates the
overall factor $(2\pi)^4 \delta(p_1+ \ldots + p_n)$, which enforces
momentum conservation at each vertex. Consequently, the momentum
$p_n = -(p_1 + \ldots + p_{n-1})$ is implicitly understood on the
left-hand side of \cref{eq:7}.

Numerical or analytical solutions of the flow equation require a
systematic truncation scheme for the effective action. A widely used
scheme is the derivative expansion, in which all correlation
functions are retained while their momentum dependence is expanded in
powers of external momenta. At leading order, corresponding to the
zeroth-order derivative expansion also known as \gls{LPA}, 
only point-like interactions without momentum
dependence are included. These contributions are encoded in a
scale-dependent effective potential that resums local interactions to
all orders in the fields.

\subsection{Quark-meson flows with diquarks}
\label{sec:quark-meson-flows}

As the initial condition for the Wetterich flow \cref{eq:WetterichEq}
at $k=\Lambda$, we take the \gls{QMD} Lagrangian introduced above.
At leading order of the derivative expansion, i.e., in \gls{LPA},
the effective action at finite quark chemical potential $\mu$ and inverse
temperature $\beta=1/T$ reads
\begin{align}
  \label{eq:qmdv-truncation}
  \nonumber &
  \Gamma_{k=\Lambda} [\Phi] = \int_0 ^\beta \!\!\!dx_0 \!\int d^3 x
  \Big\{ \bar{q} \left[ \slashed \partial - \mu \gamma_4 + g_\phi
      \left( \sigma + i \gamma_5\vec{\pi} \cdot \vec{\tau} \right)
    \right] q \\[1ex]
    \nonumber & \ \qquad\qquad + \frac{1}{2} g_\Delta \left(
      \Delta_\da \, \bar{q} \gamma_5 \tau_2 \, i\epsilon_{\da} C \bar{q}^\tp
      - \Delta_\da^* \, q^\tp C \gamma_5\tau_2 \, i\epsilon_{\da} q
    \right)                                       \\[1ex]
    & \ \qquad \qquad +
    \frac{1}{2}(\partial_\mu
    {\phi})^2 +
    U_k({\phi}^2,|\Delta|^2)-
  h\sigma \Big\} \; ,
\end{align}
with the superfield
\begin{equation}
  \label{eq:17}
  \Phi = (q, \bar{q}, {\phi},\Delta , \Delta^*)\ .
\end{equation}

The quarks $\bar{q}$ and $q$ carry two flavor degrees of freedom and
are coupled via $g_{\phi}$ to the chiral scalar $\sigma$ field and to
the pseudoscalar pion fields $\vec{\pi}$.  At lower energy scales,
emergent bound-state degrees of freedom are accounted for through a
scale-dependent effective potential $U_k({\phi}^2,|\Delta|^2)$ that
resums mesonic and diquark (self-)\-interactions to all orders in the
fields.  The scale dependence is encoded solely in the effective
potential $U_k( \phi^2, |\Delta|^2)$.

The omission of a diquark kinetic term in \cref{eq:qmdv-truncation}
corresponds to neglecting the diquark fluctuations and can be
justified in the symmetry-broken phase. In a first-principles
treatment, the formation of a diquark condensate induces gluonic
masses via the Higgs mechanism. The corresponding would-be Goldstone
modes associated with the diquark condensate are generally gauge
dependent but are completely removed from the spectrum in unitary
gauge. If, in addition, both the massive gluons and the radial
(non-Goldstone) diquark mode are assumed to be heavy, they decouple
from the low-energy dynamics and the remaining description effectively
reduces to a mean-field treatment of the diquark sector.  This
reasoning, however, does not extend to the symmetric phase, where
diquark fluctuations are expected to become dynamically relevant,
especially in the vicinity of a phase transition.

Under this assumption, the flow of the effective potential is obtained
by evaluating the Wetterich equation \cref{eq:WetterichEq} on
homogeneous constant background fields.  Exploiting the global
symmetries of the model (see \cref{sec:symmetries-model}), the chiral
field can be oriented without loss of generality in the
$\sigma$-direction,
\begin{equation} \label{eq:const_phi}
  \phi(x) = (\sigma, 0, 0, 0)
  \quad \text{with} \quad
  \sigma \in \mathbb{R} \; ,
\end{equation}
excluding a vacuum expectation value for the pseudoscalar isotriplet
components.

Similarly, the diquark field can be rotated into a fixed color
direction, which we choose as
\begin{equation} \label{eq:const_Delta}
  \Delta_\da(x) = \Delta \delta_{\da3}
  \quad \text{with} \quad
  \Delta \in \mathbb{R} \; .
\end{equation}
Due to the $SU(3)_c$ symmetry, only the modulus of the complex-valued
diquark condensate, $|\Delta| = \sqrt{\Delta^{*} \Delta}$, is
physically relevant, while its orientation in color space is
arbitrary.

The derivation of the flow, \cref{eq:WetterichEq}, requires the
regularized propagators of quarks, diquarks and mesons in this background. The bosonic contribution from the chiral sector follows
standard lines and can be found in the literature, see
e.g. \ccite{Schaefer:2004en}. The fermionic contribution is more
involved due to the Yukawa coupling between quarks and composite
diquarks, which induces a non-trivial mixing structure in the quark
propagator.

The inverse regularized quark propagator is a ($2 \times 2$)-matrix in
the field space $\Psi = (q, \bar{q})$
\begin{equation}
  \label{eq:inverse-quark-prop}
  G_{q,k}^{-1}(p) = \Gamma_{\Psi^{\tp}\Psi,k}^{(2)}(p) + R_k(p) \; .
\end{equation}
In the infrared limit $k\to 0$, the regularized propagator becomes the
full quark propagator $G_{q,k=0}$, which is related to the connected
two-point correlation functions via
\begin{equation}
  \begin{aligned}
    &
    \begin{pmatrix}
      \langle q^\tp(-p) q(l) \rangle & \langle
      \bar{q}(-p) q(l) \rangle            \\[1ex]
      \langle q^\tp(-p) \bar{q}^\tp(l) \rangle & \langle \bar{q}(-p)
      \bar{q}^\tp(l) \rangle
    \end{pmatrix} = \\ & \qquad \qquad \qquad \qquad
    G_{q,k=0}(p) \times (2\pi)^4 \delta^{(4)}(p+l) \; ,
  \end{aligned}
\end{equation}
where we use a Fourier convention in which both $q$ and $\bar{q}$ are assigned
incoming momenta.

The four-dimensional Dirac $\delta$-function ensures translational
invariance, and $\langle\ldots\rangle$ denote connected correlators.
In the present representation, the off-diagonal components describe
normal quark and antiquark propagation, whereas the diagonal entries
correspond to anomalous propagators familiar from \gls{BCS} theory. This
assignment is opposite to the conventional Nambu--\gorkov basis, in
which the anomalous contributions reside in the off-diagonal elements
\cite{Gorkov:1959bjaa, Nambu:1960tm}.

The regulator function $R_k(p)$ depends only on spatial momentum
$\vec{p}$
\begin{equation}
  R_{q,k}(p) =
  \begin{pmatrix}
    0                                            & i
    \slashed{\vec{p}}^\tp \, r_q(x) \\[1ex]
    i\slashed{\vec{p}} \, r_q(x) & 0
  \end{pmatrix} \; ,
\end{equation}
where $\slashed{\vec{p}} = p_i \gamma_i$ and $r_q(x)$ is a
dimensionless shape function of the momentum ratio $x=\sqvec{p}/k^2$. In
combination with the kinetic quark term in
$\Gamma^{(2)}_{\Psi^{\tp}\Psi,k}$, this choice amounts to a multiplicative modification of the spatial momentum,
hereafter denoted the regularized momentum $\vec{p}_{\text{reg},q} =
\vec{p} \, (1+r_q(x))$. An analogous construction applies in the bosonic
sector, yielding $\sqvec{p}_{\text{reg},b} = \sqvec{p}(1+r_b(x))$; see
\cref{app:regulator-functions} for details.

The explicit inversion of the quark propagator $G^{-1}_{q,k}$ in
\cref{eq:inverse-quark-prop}, including its non-trivial color
structure, is deferred to \cref{app:propagator_inversion}. There, we
show that the resulting propagator naturally decomposes into two
sectors, which can be expressed in terms of the color-space projectors
\begin{equation}
  \label{eq:67}
  P_{rg} = \text{diag} (1,1,0)\ ,\quad \text{and} \quad P_b =
  \text{diag} (0,0,1)\ ,
\end{equation}
corresponding to the paired red-green and the spectator blue sector,
respectively.  The full propagator then reads
\begin{align}
  \label{eq:nambu_gorkov_propagator}
  G_{q,k} (p) = &
  \begin{pmatrix}
    \Xi^\dagger (p)        & G_\Delta (p)\\[1ex]
    \bar{G}_\Delta^\tp (p)& \Xi (p)
  \end{pmatrix} P_{rg}
  +
  \begin{pmatrix}
    0             & G_0 (p)\\[1ex]
    \bar{G}_0^\tp (p) & 0
  \end{pmatrix} P_{b} \; .
\end{align}
In the following, for notational simplicity, we denote the regularized momenta
$\vec{p}_{\text{reg},q}$ by $p$ in all 
propagators and energy projectors.

One immediately recognizes that consistently with the symmetry-breaking
pattern $SU(3)_c \to SU(2)_c$, the two quark colors in the red-green
sector remain degenerate, while the blue quark propagates as an
unpaired spectator mode.

At finite temperature, we employ the standard imaginary-time (Matsubara)
formalism, in which the Euclidean time direction is compactified and as a
consequence, the temporal momentum component becomes discrete
\cite{Bellac:2011kqa}. 
For antisymmetric fields the components are replaced by fermionic
Matsubara frequencies $\nu_n = (2n+1) \pi T$.

In the blue sector, at finite temperature and chemical potential, the
propagator matrix elements decompose into contributions associated
with positive- and negative-energy modes and take the form \
\begin{equation}
  \label{eq:72}
  G_0 (p)      =
  \frac{1}{i \nu_n - \epsilon_q^-} P_+ \gamma_4
  + \frac{1}{i \nu_n + \epsilon_q^+} P_- \gamma_4 \; .
\end{equation}
The operators $P_{\pm}$ and $\bar{P}_{\pm}$ denote the Dirac energy
projectors and are defined in \cref{eq:P,eq:Pbar}, respectively. The
charge-conjugate propagator satisfies
\begin{equation}
  \label{eq:73}
  \bar{G}_0 (p) = -G_0 (-p)\ .
\end{equation}

This representation of the propagator makes the pole structure
manifest and directly yields the finite-density dispersion relations
\begin{align}
  \label{eq:13}
  \begin{split}
    \epsilon_q^\pm =  \epsilon_q  \pm \mu  
    \qquad 
    \text{with} \qquad \epsilon_q (\vec{p})  =  \sqrt{ \sqvec{p} + m_q^2}\ .
  \end{split}
\end{align}
The quark mass $m_q = g_{\phi} \sigma$ is dynamically generated by the
sigma field $\sigma$.

In the red-green sector, the propagator matrix elements take a more
involved form. They read
\begin{equation}
\label{eq:74}
  G_\Delta (p) =
  \frac{i \nu_n + \epsilon_q^-}{(i\nu_n)^2 - E_q^{-2}} P_+ \gamma_4
  + \frac{i \nu_n - \epsilon_q^+}{(i\nu_n)^2 - E_q^{+2}} P_- \gamma_4 \; , \\
\end{equation}
with anomalous propagators 
\begin{equation}
\label{eq:75}
  \Xi (p)   = - \Delta_{\text{gap}} \tau_2 i\epsilon_3 C \gamma_5 \left(
    \frac{ \bar{P}_+}{(i\nu_n)^2 - E_q^{+2}}
    + \frac{  \bar{P}_-}{(i\nu_n)^2 - E_q^{-2}}
  \right) \; ,
\end{equation}
and correspondingly for $\Xi^{\dagger} (p)$. The
charge-conjugate propagator $\bar{G}_\Delta$ is
related to $G_\Delta$ analogously to the blue sector and satisfies
\begin{equation}
  \bar{G}_\Delta(p) = -G_\Delta(-p) \; .
\end{equation}

The quasiparticle dispersion relations are given by
\begin{equation}
  E_q^\pm = \sqrt{\epsilon_q^{\pm 2} + \Delta_{\text{gap}}^2
  } \quad \text{with }\quad\Delta_{\text{gap}} = g_\Delta \Delta\; .
\end{equation}
These relations show that, in the presence of a finite diquark
condensate, the red and green quarks acquire a gap at the
Fermi-surface, characteristic of a superconducting phase. The
off-diagonal propagators $G_\Delta (p)$ and $\bar{G}_\Delta^{\tp} (p)$
describe the propagation of gapped red and green quarks. By contrast,
the diagonal components $\Xi (p)$ and $\Xi^\dagger (p)$ are
non-vanishing only for a finite diquark condensate and are directly
associated with the spontaneous breaking of the $U(1)_B$ symmetry.

With the full propagator at hand, the Wetterich equation can be
employed to derive the flow of the effective potential
$U_k$. Performing the trace over color, flavor, and Dirac indices and
evaluating the Matsubara sums yields the general flow
equation\footnote{For brevity, temperature and chemical potential
  arguments are omitted hereafter.}
\begin{align}
  \label{eq:flow_arbitrary_shape}
  \partial_t  U_k (\sigma,\Delta ) & = \frac{1}{4} \int_{\vec{p}}
  \sqvec{p} (\partial_t r_b)
  \bigg\{
    \frac{3}{\epsilon_\pi}\coth \frac{\epsilon_\pi}{2T}
    + \frac{1}{\epsilon_\sigma}\coth \frac{\epsilon_\sigma}{2T}  \nonumber\\[1ex]
    & \quad - \frac{2N_f}{\epsilon_q} \left[ \tanh \left(
      \frac{\epsilon^+_q }{2T} \right) + \tanh \left(
      \frac{\epsilon^-_q }{2T} \right) \right.                    \\[1ex]
     +\, 2 & \left. \frac{\epsilon^+_q }{E_q^+} \tanh \left(
      \frac{E_q^+}{2T} \right) + 2\frac{\epsilon^-_q }{E_q^-} \tanh
    \left( \frac{E_q^-}{2T} \right) \right]
  \bigg\} \; , \nonumber
\end{align}
where unified regulator shape functions have been employed, see
\cref{eq:37}.

Using the shorthand notation for derivatives with respect to
the squared fields
\begin{equation}
  U_k^{(n,m)} (\sigma, \Delta) = \frac{\partial^{n+m} U_k (\sigma, \Delta)}
  {\partial(\sigma^2)^n \partial (\Delta^2)^m} \; .
\end{equation}
the (pseudo-)scalar dispersion relations  can
be expressed as
\begin{equation}
  \label{eq:22}
  \epsilon_i =
  \sqrt{\sqvec{p}_{\text{reg},b}  + m^2_i}\ , \quad i = \sigma, \pi \; ,
\end{equation}
with meson masses
\begin{align}
  \label{eq:29}
  & m_\pi^2  = 2 U^{(1,0)}_k (\sigma,\Delta) \; , \\[1ex]
  \label{eq:sigma_mass}
  \text{and} \quad & m_\sigma^2  = 2 U^{(1,0)}_k (\sigma,\Delta)
  + 4 \sigma^2 U^{(2,0)}_k (\sigma,\Delta) \; ,
\end{align}
which depend on the diquark gap $\Delta$ only through the effective
potential.

\begin{figure}[!t]
  \centering
  \includegraphics{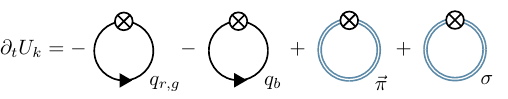}
  \caption{\label{fig:diagramflowequation} Diagrammatic flow of the
    quark-meson-diquark model in \gls{mLPA}. Solid lines denote dressed
    quark propagators for the red and green quarks $q_{r,g}$
    (participating in the superconducting phase) and the blue quark
    $q_b$ (non-participating). Double solid lines indicate the dressed
    $\sigma$ and $\vec{\pi}$ propagators. Crosses mark regulator
  insertions.}
\end{figure}

Using a flat regulator (\cref{app:regulator-functions}) and replacing
the regularized momenta by the \gls{RG} scale $k$,
$\sqvec{p}_{\text{reg},b/q} \rightarrow k^2$, the flow equation for
the effective potential reads
\begin{align}
  \label{eq:full_flow}
  \partial_t U_k(\sigma, \Delta) = & \frac{k^5}{12 \pi^2} \bigg\{
    \frac{3}{\epsilon_\pi}\coth \frac{\epsilon_\pi}{2T}
    + \frac{1}{\epsilon_\sigma}\coth \frac{\epsilon_\sigma}{2T}
    \nonumber\\[1ex]
    & \quad - \frac{2N_f}{\epsilon_q} \bigg[ \tanh \frac{ \epsilon_q^{+}
      }{2T} + \tanh
      \frac{\epsilon_q^{-}}{2T}
      \\[1ex]
      & \qquad + 2\frac{\epsilon_q^{+} }{E_q^+} \tanh \frac{E_q^+}{2T} +
    2\frac{\epsilon_q^{-} }{E_q^-} \tanh\frac{E_q^-}{2T} \bigg]
  \bigg\}\ . \nonumber
\end{align}
This flow receives contributions from four distinct threshold
functions, which can be represented as loops with regulator insertions
(crosses), as illustrated in \cref{fig:diagramflowequation}. The
doubled lines correspond to the $\sigma$ and $\vec{\pi}$ meson loops,
(first line of \cref{eq:full_flow}), while the second line with quark
and antiquark contributions accounts for the unpaired ("blue") quark
loop with fermion sign, labeled $q_b$ in the figure. Finally, the
coupled red-green quark loop, $q_{r,g}$, carries the diquark gap
$\Delta$.

In the following, we refer to this as the \gls{mLPA}, 
to distinguish it from the full \gls{LPA}
including diquark fluctuations.  The thermodynamic (grand) potential is then given by the infrared solution of the flow equation:
\begin{equation}
  \label{eq:qmv_potential}
  \begin{aligned}
    \Omega(\sigma,\Delta) = U_{k=0}(\sigma, \Delta) - h\sigma\ .
  \end{aligned}
\end{equation}

For a given temperature $T$ and chemical potential $\mu$, the
thermodynamic phases are determined by the gap equations (the equation
of motion)
\begin{equation}
  \label{eq:full_gap_equations}
  \left. \frac{\partial \Omega (\sigma, \Delta)}{\partial \sigma}
  \right|_{\bar\sigma, \bar\Delta}
  = \left. \frac{\partial \Omega (\sigma, \Delta)}{\partial \Delta}
  \right|_{\bar\sigma, \bar\Delta}
  = 0 \; ,
\end{equation}
which minimize the grand potential at the spacetime-independent
condensates $\bar\sigma$ and $\bar\Delta$. Generally, thermodynamic
quantities evaluated on the equation of motion are denoted by a bar,
as for the physical quark mass $\bar m_q$ and physical diquark
gap $\bar \Delta_{\text{gap}}$.

All thermodynamic quantities follow from standard derivatives of
$\Omega (\sigma, \Delta)$, e.g., the entropy density is
\begin{equation}
  \label{eq:30}
  s = - \frac{\partial \Omega(\bar\sigma,\bar\Delta)}{\partial T} \; .
\end{equation}

It is instructive to consider certain limits of the flow equation to
gain insight into its structure.

\subsubsection{Zero temperature flows}

At zero temperature and finite chemical potential, the flow \cref{eq:full_flow}
simplifies to
\begin{equation}
  \begin{aligned}
    & \partial_t U_k(\sigma, \Delta) = \frac{k^5}{12 \pi^2}
    \bigg\{
      \frac{3}{\epsilon_\pi}
      + \frac{1}{\epsilon_\sigma}                              \\
      & \quad - \frac{4 N_f}{\epsilon_q} \left[
        \theta(\epsilon_q^-)
        + \frac{\epsilon_q^+}{E_q^+}
        + \frac{\epsilon_q^-}{E_q^-}
      \right]
    \bigg\}\ .
  \end{aligned}
\end{equation}
For $\Delta= 0$, the fermionic energy ratios 
reduce to
\begin{equation}
  \label{eq:15}
  \frac{\epsilon^+_q}{E^+_q} \to 1\qquad \text{and} \qquad
  \frac{\epsilon^-_q}{E^-_q} \to 2 \theta (\epsilon^-_q) -1\ ,
\end{equation}
so that the fermionic flow $N_c \theta(\epsilon^-_q)$ exhibits the
Silver-Blaze property at $T=0$: fermions and antifermions contribute
only when $\epsilon^-_q>0$, i.e., for $\epsilon_q > \mu$, see
\cref{sec:silver-blaze-rule}.

\subsubsection{Quark-meson flows}

For $\Delta=0$ and finite temperature, the
quasi-particle energies reduce to
\begin{equation}
  \label{eq:11}
  E_q^+ \to \epsilon_q^+ \qquad \text{and} \qquad E_q^- \to
  |\epsilon_q^- |\ .
\end{equation}
Consequently, the flow converges to that of a quark-meson flow with $N_c=3$
degenerate colors:
\begin{equation}
  \label{eq:qmv_flow}
  \begin{aligned}
    \partial_t U_k(\sigma, \Delta=0) =\frac{k^5}{12 \pi^2} \left\{ -
      \frac{2N_f N_c}{\epsilon_q} \bigg[
        \tanh \frac{ \epsilon_q^{+} }{2T} + \right. \\
      \left. \tanh \frac{\epsilon_q^{-}}{2T} \bigg] +
      \frac{3}{\epsilon_\pi}\coth \frac{\epsilon_\pi}{2T} +
      \frac{1}{\epsilon_\sigma}\coth \frac{\epsilon_\sigma}{2T}
    \right\}\ .
  \end{aligned}
\end{equation}

\subsection{Parameter fixing}
\label{sec:param-fixing}

To solve the flow equation \cref{eq:full_flow} numerically, the
required parameters are fixed from vacuum physics.  While the
$\sigma$-field acquires a non-vanishing vacuum expectation value due
to the spontaneous chiral symmetry breaking, the diquark fields are
expected to develop nonzero mean values only in dense
matter. Accordingly, all couplings and masses in the
scalar-pseudoscalar interaction channel of \cref{eq:full_flow} can be
fixed in the standard way from vacuum observables, in contrast to the
diquark sector. We adopt $k=\Lambda = 600$ MeV\footnote{Later, in the
  context of \gls{RGC}, see \cref{sec:renorm-group-cons}, we
  denote this \gls{UV} scale at which the initial conditions for the
  effective action are specified by $\Lambda'$.}  as a typical \gls{UV}
scale, in line with recent \gls{QCD}-motivated low-energy model calculations
see e.g. \ccites{Fu:2021oaw, Fu:2023lcm}.

The two parameters $a_1$ and $a_2$ of the \gls{UV} potential, cf.~\cref{eq:2},
\begin{equation}
  \label{eq:6}
  U_{k=\Lambda'}(\sigma,\Delta)
  = a_1 \sigma^2 + a_2 \sigma^4 + b_1 \Delta^2 \; ,
\end{equation}
are fixed to the chiral condensate $\bar{\sigma} = f_\pi = 92.4$ MeV
and the \gls{IR} sigma mass
$m_{\sigma} = 560 \MeV$ in the vacuum\footnote{ For numerical
  convenience, we work at small but finite temperature, $T=1$ MeV,
  rather than at strictly zero temperature.}.  
The scalar-pseudoscalar Yukawa coupling is set to $g_\phi = 3.25$ which
corresponds to a vacuum quark mass $m_q = 300 \MeV$. Finally, we adjust the
explicit chiral symmetry breaking parameter $h$ such that the \gls{IR} pion mass
$m_{\pi} = 137 \MeV$ is reproduced. 

The medium contribution proportional to $\mu^2 \Delta^2$, originating from the
diquark kinetic term in the classical action \cref{eq:classical_action}, can be
neglected at the \gls{UV} scale. This corresponds to a vanishing diquark
wave-function renormalization at that scale, consistent with the interpretation
of the diquark as a composite field introduced via a Hubbard-Stratonovich
transformation without an associated kinetic term.

The remaining diquark Yukawa coupling $g_\Delta$ is treated as a free
parameter. Rather than performing an extensive parameter scan, we
choose a few representative values to illustrate the generic physical
behavior. 
The squared diquark mass parameter $b_1$ is fixed at the \gls{UV}
  scale such that the resulting infrared curvature mass,
\begin{equation}
  \label{eq:9}
  m^2_{\Delta,\text{curv}} \equiv
  \Omega^{(0,1)} (\sigma,\Delta) |_{\text{gap}} \; ,
\end{equation}
evaluated on the equation of motion, \cref{eq:full_gap_equations}, is
of the order of the expected physical scale.
  Consistent with the \gls{MFA} and \gls{mLPA} approximations, we
  neglect the diquark wave-function renormalization in the definition
  of the curvature mass.  We return to this approximation at the end
  of \cref{sec:pole_and_curv_mass}. Corresponding estimates for the
diquark pole mass are given in Refs.~\cite{Hess:1998sd, Oettel:2000jj,
  Maris:2002yu}. Specifically, we set the curvature mass to twice the
constituent quark mass, $m_{\Delta,\text{curv}} = 600 \MeV$.
  This choice is motivated by the assumption that, in vacuum, the
  diquark curvature mass approximately coincides with the
  corresponding pole mass, although this relation is not guaranteed to
  hold; see \cref{sec:pole_and_curv_mass}. A detailed analysis of the
  relation between diquark pole and curvature masses in vacuum
  \gls{QCD} will be presented elsewhere \cite{mire2026prep}.  Since
  the present work focuses on qualitative features of the phase
  structure, the results are insensitive to the precise parameter
  values. A summary of all parameters is given in
  \cref{app:uv-param-numer}.

\subsection{Recovering mean-field flows}
\label{sec:recov-mean-field}

It is instructive to consider the mean-field approximation as a
reference limit, which allows us to isolate and quantify the impact of
bosonic fluctuation effects included in the functional approach.  In
the \gls{MFA}, quantum and thermal fluctuations of the bosonic fields
are neglected in the path integral representation of the grand
potential. The mesonic quantum fields are replaced by their classical
expectation values, and only the fermionic (quark) determinant is
retained, corresponding to the integration over the quark loop. The
standard \gls{MFA} can also be recovered directly from our quark-meson
flow equation including diquarks, \cref{eq:full_flow}, by neglecting
the bosonic contributions.  For an arbitrary regulator, this
corresponds to discarding the pion and sigma terms in
\cref{eq:flow_arbitrary_shape}. This \gls{MFA} flow can be rewritten
in the compact form
\begin{equation}
  \label{eq:19}
  \partial_t U^{\text{MFA}}_k(\sigma,\Delta) = \partial_t f_k(\sigma,
  \Delta) \; ,
\end{equation}
with the auxiliary function
\begin{align}
  \label{eq:14}
  f_k&(\sigma,\Delta) = -2N_f \!\!\int_{\vec{p}} \bigg\{
    \frac{1}{2}\!\left( |\epsilon_q^-| + 2E_q^-  \right)+
    \frac{1}{2}\!\left( |\epsilon_q^+| + 2E_q^+ \right) \nonumber\\
    & \qquad +\!T \ln \left( 1 + e^{-\beta |\epsilon_q^-| } \right)
    + 2T \ln \left( 1 + e^{-\beta E_q^-} \right) \\
    &  \qquad + \!T \ln \left( 1 + e^{-\beta |\epsilon_q^+|} \right)
  + 2T \ln \left( 1 + e^{-\beta E_q^+}\right)\!\!\bigg\}  \nonumber
  \ .
\end{align}
Here, we have used the relations
\begin{equation}
  \label{eq:3}
  \partial_t \epsilon_q= \frac{\sqvec{p}(\partial_t r_b)}{2\epsilon_q} \; ,
  \quad \quad
  \partial_t E_q^\pm = \frac{\epsilon_q^\pm}{E_q^\pm}
  \partial_t \epsilon_q \; ,
\end{equation}
in conjunction with \cref{eq:37},
as well as the identity
\begin{equation}
  \label{eq:8}
  (\partial_t f) \tanh\frac{f}{2T} =
  \partial_t \left[f + 2T \ln\left(
      1 + e^{-\beta f}
  \right)\right]\ ,
\end{equation}
which holds for an arbitrary function $f(t)$. In \cref{eq:14} one can
clearly see how the contributions from fermions and antifermions combine to account for all three color degrees of freedom.

Since the mean-field flow \cref{eq:19} is both \gls{IR} and
\gls{UV} finite, the integration over the \gls{RG} scale can be
carried out analytically
\begin{equation}
  \label{eq:21}
  \int_{\Lambda}^0 dk \partial_k
  U^{\text{MFA}}_k (\sigma, \Delta) = f_0(\sigma, \Delta) -
  f_{\Lambda}(\sigma,\Delta)\ .
\end{equation}
The grand potential in \gls{MFA}, $\Omega^{\text{MFA}}$, is obtained by
supplementing the chirally symmetric effective potential in the
infrared $U_{k=0}^{\text{MFA}}$ with the explicit chiral symmetry
breaking term
\begin{align}
  \begin{split}
    \label{eq:scale-dependent-mean-field-potential}
    \Omega^{\text{MFA}} (\sigma,\Delta) = & \ U^{\text{MFA}}_{k=0} - h
    \sigma \\[1ex]
    = & \ U^{\text{MFA}}_{\Lambda}
    (\sigma, \Delta) -h\sigma \\[1ex]
    &+ f_0(\sigma, \Delta) - f_{\Lambda}(\sigma,\Delta) \; .
  \end{split}
\end{align}
Due to the finiteness of the flow equation, the function
$f_{k=\Lambda}$ effectively controls the divergent momentum integral
appearing in $f_{k=0}$.  Different choices of shape functions $r_q(x)$
correspond to different regularization schemes in the standard
mean-field formulation. For instance, selecting the sharp shape
function
\begin{equation}
  \label{eq:12}
  (1 + r_q(x))^2 = \frac{1}{\theta(x-1)} \; ,
\end{equation}
amounts to imposing a conventional three-momentum cutoff
$|\vec{p}\,|=\Lambda$. In this case, the resulting
mean-field potential coincides with the standard expressions
commonly used in QM-model studies, see, e.g.,~\ccite{Schaefer:2011ex}.

Note that the \gls{MFA} of the grand potential can also be obtained by
standard regularization and renormalization of the
\gls{UV}-divergent\footnote{Including diquark degrees of freedom
  introduce additional $\mu$-dependent divergences.} contribution from
the one-loop quark determinant to the effective potential, see
e.g.~\ccites{Skokov:2010sf,Andersen:2024qus,Gholami:2025afm}. Performing
the \gls{MFA} within the \gls{FRG} framework, however, ensures that
the same nonperturbative regularization and renormalization schemes
are used when comparing with the \gls{mLPA} results.

\subsection{Renormalization group consistency}
\label{sec:renorm-group-cons}

In order to remove cutoff artifacts in our low-energy effective theory, we
employ the concept of renormalization-group consistency (\gls{RGC}), see
\ccites{Pawlowski:2005xe, Pawlowski:2015mlf, Braun:2018svj}. Here, we only
briefly review the general concept of \gls{RGC} and its treatment of medium
divergences. For a more comprehensive discussion in the context of the \gls{QMD}
model, we refer the reader to \ccite{Gholami:2025afm}. \gls{RGC} requires a
consistent regularization and renormalization procedure such that the explicit
cutoff dependence of the bare action $\Gamma_{\Lambda}$ in the \gls{UV}
compensates the cutoff dependence generated along the flow towards the infrared.
As a consequence, the full quantum effective action becomes cutoff independent,
\begin{equation}
  \label{eq:61}
  \Lambda \frac{\mathrm{d}\Gamma}{\mathrm{d}\Lambda} = 0 \ .
\end{equation}
In particular, it has been applied to the computation of the \gls{EoS} at finite
temperature \cite{Herbst:2013ufa} and shown to remove cutoff artifacts in the
presence of a diquark condensate \cite{Braun:2018svj, Gholami:2024diy}.

However, in practice, additional external scales of the theory, such
as the temperature $T$ or the quark chemical potential $\mu$, can
violate this independence, so that the bare action $\Gamma_{\Lambda}$
must vary with changes in these external scales to ensure \gls{RGC}.
In the following, we outline how cutoff artefacts can be consistently
removed in our \gls{QMD} model by enforcing \gls{RGC}.

The idea is to formally integrate the vacuum flow upward to some
higher scale $k=\Lambda>\Lambda'$, assuming that the initial
conditions are originally specified by the effective action at a fixed
scale $\Lambda'$
\begin{equation}
  \label{eq:general-rgc-initial-condition}
  \Gamma_\Lambda[\Phi] = \Gamma_{\Lambda'}[\Phi]
  + \int_{\Lambda'}^\Lambda \frac{dk'}{k'} \mathcal{F}_{k'}[\Phi] \; ,
\end{equation}
with the flow $\partial_t \Gamma_k[\Phi] = \mathcal{F}_k[\Phi]$.  In
this way, a modified initial condition for the effective action at the
higher scale $\Lambda$ is obtained which yields the same effective
action $\Gamma[\Phi]$ in the limit $k \to 0$.  For any intermediate
scale $k \neq \Lambda$, \gls{RGC} then follows directly:
\begin{align} \label{eq:16}
  \nonumber
    \partial_\Lambda \Gamma_k[\Phi] &=
    \partial_\Lambda \int^\Lambda_{\Lambda'} \frac{dk'}{k'}
    \mathcal{F}_{k'}[\Phi]
    + \partial_\Lambda \int_\Lambda^k \frac{dk'}{k'} \mathcal{F}_{k'}[\Phi]\\[1ex]
    &= \partial_{\Lambda} \Gamma_{\Lambda}[\Phi] - \mathcal{F}_{\Lambda}
    [\Phi] = 0 \; ,
\end{align}
demonstrating that the resulting effective action is independent of
the choice of $\Lambda$.

\subsubsection{RG consistency in the mean-field flow}
\label{sec:rg-cons-with}

The implementation of \gls{RGC} at the mean-field level of a \gls{QMD}
model is straightforward. In principle, the corresponding \gls{RGC}
initial condition follows directly from the definition of the
scale-dependent mean-field potential in \cref{eq:19}.

However, one important difference is that the initial potential at the
\gls{UV} scale -- denoted henceforth by the scale $\Lambda'$ --
contains an explicit \gls{UV}-divergent term proportional to $\mu^2$
in \cref{eq:14}.  This medium divergence grows without bound as
$\Lambda \to \infty$, see also \ccite{Gholami:2025afm} for
details.\footnote{Due to the presence of a Landau pole, the limit
    $\Lambda \to \infty$ cannot be taken. Accordingly, \gls{RGC}
    ensures \gls{UV}-cutoff independence only below the cutoff scale set by
    the Landau pole.} To handle this divergence without introducing
an explicit diquark wave-function renormalization, we perform a Taylor
expansion of the upward flow in powers of $\mu$ and add the resulting
contributions to the right-hand side of
\cref{eq:general-rgc-initial-condition}. The first term, linear in
$\mu$, vanishes due to the symmetry of the effective potential under
$\mu \to -\mu$, corresponding to particle-antiparticle interchange.
The second term in the \gls{RGC} construction, however, consistently
captures the medium divergence.

The $\mu$-dependent \gls{RGC}-improved mean-field 
condition at $k=\Lambda$ then reads, reinstating the explicit $T$ and
$\mu$ dependence,
\begin{align}
  \label{eq:66}
  \nonumber
  U_{\Lambda}^{\text{MFA}}(\sigma,\Delta; &\, \mu)=
  U_{\Lambda'}^{\text{MFA}}(\sigma,\Delta; 0,0) + \mathcal{F}_{\Lambda'
  \to \Lambda}(\sigma,\Delta; 0, 0) \\ &  +
  \frac{\mu^2}{2} \left. \partial_\mu^2 \right|_{\mu=0}
    \mathcal{F}_{\Lambda'\to\Lambda} \left(\sigma,\Delta;0,\mu
  \right) \; ,
\end{align}
with the abbreviation for the integrated mean-field flow, see
\cref{eq:19},
\begin{align}
  \label{eq:65}
  \mathcal{F}_{k_1 \to k_2}(\sigma,\Delta; T, \mu) &=
  \int_{k_1}^{k_2} \frac{dk'}{k'}\, \partial_t
  f_{k'}(\sigma,\Delta; T,\mu) \\
  & =  f_{k_2}(\sigma,\Delta; T,\mu) - f_{k_1}(\sigma,\Delta; T,\mu)
  \; . \nonumber
\end{align}

At the higher scale $\Lambda$, no temperature-dependent medium
divergences arise within the \gls{MFA}, such that the mean-field potential
$U^{\text{MFA}}_{\Lambda}$ is temperature-independent. This contrasts
with the initial potential at the lower scale $\Lambda'$, which is
obtained by integrating the fermionic flow from $k=\Lambda$ to
$\Lambda'$:
\begin{align} \label{eq:41}
  \nonumber
  U^{\text{MFA}}_{\text{init},\Lambda'}(\sigma,\Delta; T,\mu) = &
  U_{\Lambda}^{\text{MFA}}(\sigma,\Delta; \mu) \\ & \qquad
  + \mathcal{F}_{\Lambda \to \Lambda'}(\sigma,\Delta;T,\mu) \; .
\end{align}
Eventually, the scale-dependent mean-field potential follows by
integrating the fermionic flow from $k=\Lambda'$ with the modified
initial condition \cref{eq:41}:
\begin{align} \label{eq:44}
  \nonumber
  U_k^{\text{MFA}}(\sigma,\Delta;T,\mu) = &
  U_{\text{init},\Lambda'}^{\text{MFA}}(\sigma,\Delta;\mu) \\ & \qquad
  + \mathcal{F}_{\Lambda' \to k}(\sigma,\Delta;T,\mu)\; .
\end{align}

\ccite{Gholami:2025afm} demonstrated that the subtraction of
  the medium divergence as in \cref{eq:66} is not unique. In
  particular, the \gls{RGC} scheme employed here (the $\sigma\Delta$
  scheme in \ccite{Gholami:2025afm}) yields larger diquark gaps
  compared to the other schemes and does not satisfy the \gls{BCS} relation
  for large chemical potentials. Nevertheless, the qualitative
  behavior of the \gls{QMD} model was found to be largely insensitive to the
  choice of the \gls{RGC} scheme, which suffices for the present
  analysis.

\subsubsection{RG consistency in the mesonic LPA flow}
\label{sec:rg-consistency-lpa}

Implementing \gls{RGC} within the \gls{LPA} is in general more subtle
than in the \gls{MFA}. Mathematically, \gls{RG} flows form a
semi-group rather than a group: the coarse-graining from the \gls{UV}
to the \gls{IR} is irreversible and no inverse transformation
exists. While this limitation is less severe in the \gls{MFA},
enforcing \gls{RGC} in \gls{LPA} (or beyond) requires particular care.

Technically, this originates from the diffusive character of the
bosonic sector. From a fluid-dynamical perspective, the flow
\cref{eq:full_flow} can be recast as a nonlinear advection-diffusion
equation. In this interpretation, the fermionic sector acts as a
source term, while the pion contribution induces a nonlinear advective term
and the sigma contribution generates diffusion, see
\ccites{Koenigstein:2021syz, Koenigstein:2021rxj} for details.  As in
standard diffusion processes in fluid dynamics, such equations define
an irreversible evolution.  Consequently, upward integration cannot be
employed to reconstruct \gls{RGC} initial conditions, since diffusion
proceeds independently of the direction in which the flow is
integrated.

To proceed, we note that in the \gls{UV} the flow is dominated by
fermionic contributions, while bosonic contributions are strongly
suppressed due to the increasing meson masses in the threshold
functions.  At asymptotically large $k$, the bosonic contributions
vanish and thus become irrelevant in this regime. We therefore can
proceed similarly to the mean-field procedure described above and integrate
only the fermionic flow upward to obtain the \gls{RGC} condition, and
include the bosonic flow once the parameter-fixing scale $\Lambda'$ is
reached.  This procedure has also been proposed in
\ccite{Braun:2018svj}.

Thus, in \gls{mLPA} the $T$- and $\mu$-dependent \gls{RGC} initial
condition at the scale $k=\Lambda'$ is given by
\begin{equation}
  \label{eq:38}
  \begin{aligned}
    & U_{\text{init},\Lambda'}(\sigma, \Delta; T, \mu) =
    U_{\Lambda'}(\sigma, \Delta;0,0) + \mathcal{F}_{\Lambda'
    \to \Lambda}(\sigma, \Delta; 0, 0) \\ & \quad +
    \frac{\mu^2}{2} \left. \partial_\mu^2 \right|_{\mu=0}
      \mathcal{F}_{\Lambda'\to\Lambda}(\sigma,\Delta;0,\mu)
     + \mathcal{F}_{\Lambda \to
    \Lambda'}(\sigma,\Delta; T, \mu) \; ,
  \end{aligned}
\end{equation}
with the integrated mean-field flow
$\mathcal{F}_{k_1\to k_2}(\sigma,\Delta; T, \mu)$, \cref{eq:65}.  

In summary, we integrate the flow upward from $\Lambda'$ to $\Lambda$
and perform the expansion in $\mu$ including only the fermionic
contributions. Subsequently, the flow is integrated back in medium,
i.e., at finite $T$ and $\mu$, again retaining only the fermionic
contributions from $\Lambda$ to $\Lambda'$. This yields the \gls{RGC}
initial conditions at the scale $k=\Lambda'$, from which the mesonic
fluctuations are subsequently included.

\section{Diquark Two-point Function}
\label{sec:corr-funct-at}

Before presenting our numerical results, we investigate the diquark
two-point function. Its pole structure encodes the existence and
masses of diquark excitations, thereby offering direct insight into bound-state formation and the onset of diquark condensation as
implied by the Silver-Blaze property.

\subsection{Silver-Blaze property}
\label{sec:silver-blaze-rule}

The Silver-Blaze property \cite{Cohen:2003kd} imposes non-trivial
constraints on the 1PI $n$-point functions and thereby provides
valuable guidance for understanding the phase structure of a theory,
in particular at finite density.  For a general discussion of the
Silver-Blaze property, see \ccite{Marko:2014hea, Khan:2015puu}; for
its phenomenological and field-theoretical implications, see
\ccite{Fu:2016tey, Braun:2020bhy}.

At vanishing temperature, it states
that the generating functional (or equivalently the partition
function) remains independent of the chemical potential $\mu$ below
the Silver-Blaze threshold $\mu_c$.  For a general 1PI $n$-point
function, cf.~\cref{eq:7}, this implies that for $\mu < \mu_c$
\begin{equation}
  \label{eq:silver_blaze_npoint}
  \Gamma_{\Phi_{1} \dots \Phi_{n}}^{(n)}(p_1,\dots,p_{n-1}; \mu)
  = \Gamma_{\Phi_{1} \dots \Phi_{n}}^{(n)}(\tilde p_1,\dots,\tilde p_{n-1}; 0) \; ,
\end{equation}
with  shifted momenta
\begin{equation}
  \tilde p_\varphi = (p_0 + i q_\varphi \mu, \vec{p}) \; ,
\end{equation}
where $q_\varphi \in \mathbb{Z}$ denotes the fermion number (or $q_{\varphi}/3$
the baryon number) of the field $\varphi$. The critical chemical potential
$\mu_c$ is determined by the smallest pole mass and the fermion number
$q_{\varphi}$ of those fields that couple to the chemical potential.

For (anti)diquark fields, the fermion numbers are
\begin{equation}
  \label{eq:diquark_charge}
  q_{\Delta} = +2
  \quad \text{and} \quad
  q_{\Delta^*} = -2 \; ,
\end{equation}
since (anti)diquarks consist of two fermions or two antifermions,
respectively. Applied to the diquark and anti-diquark two-point
functions, the Silver-Blaze property yields
\begin{equation}
  \label{eq:silver_blaze_diquark}
  \Gamma^{(2)}_{\Delta^*_\da\Delta_\da}(p; \mu)
  = \Gamma^{(2)}_{\Delta^*_\da\Delta_\da}(\tilde p; 0) \; ,
\end{equation}
with the shifted momenta $\tilde p = (p_0 \pm 2 i\mu, \vec{p})$.

\begin{figure*}[!t]
  \centering
  \includegraphics[width=0.65\textwidth]{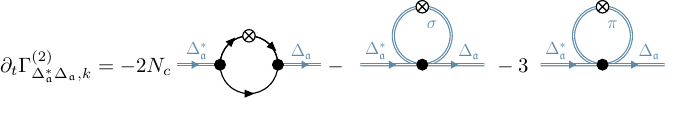}
  \caption{Vacuum flow of the diquark two-point function including
    bosonic (sigma and pion) fluctuations. The flow of the two-point
    function in \gls{RPA} is just the first diagram. 
      }
  \label{fig:two_point_flow_lpa}
\end{figure*}

This relation can be used to infer the chemical-potential dependence of the
(anti)diquark pole masses. In general, the pole mass of a (complex scalar)
particle $\phi$ is defined as the zero of its inverse propagator at vanishing
three-momentum, $\vec{p}=0$,
\begin{equation}
  \label{eq:20}
  \Gamma^{(2)}_{\phi^{*} \phi} (p_0 = im_{\phi}, 0; \mu)= 0 \; .
\end{equation}
For a diquark with fermion number $q_{\Delta} = +2$, a pole mass at
$\mu=0$ satisfies $\tilde{p}_0 = i m_{\Delta,\text{pole}}$ which
translates at finite $\mu$ to
$p_0 = i (m_{\Delta,\text{pole}} - 2\mu)$.  Similarly, for an
anti-diquark, a pole mass at negative real-time momenta
$\tilde{p}_0= -i m_{\Delta^*,\text{pole}}$ leads to
$p_0 = i (m_{\Delta^*,\text{pole}} + 2\mu)$.
Hence, the diquark pole mass \textit{decreases} linearly with $\mu$,
\begin{equation}
  m_{\Delta,\text{pole}}(\mu) = m_{\Delta,\text{pole}}(0) - 2\mu \; ,
\end{equation}
while the anti-diquark pole mass \textit{increases} with $\mu$,
\begin{equation}
  m_{\Delta^{*},\text{pole}}(\mu) = m_{\Delta^{*},\text{pole}}(0) + 2\mu \; .
\end{equation}

As a further consequence of the Silver-Blaze property for $n$-point
functions, one can directly infer the chemical-potential dependence of
the diquark curvature mass defined in \cref{eq:9}. The dependence of
the curvature mass on the chemical potential, defined through the
diquark two-point function at vanishing external momentum, is via
Silver-Blaze entirely determined by the frequency dependence of the
vacuum diquark two-point function
\begin{equation}
  m^2_{\Delta,\mathrm{curv}}(\mu) 
  \equiv \Gamma^{(2)}_{\Delta^*\Delta}(0, 0; \mu)
  =  \Gamma^{(2)}_{\Delta^*\Delta}(2i\mu , 0; 0) \; .
\end{equation}
Consequently, a vanishing diquark pole mass at the Silver-Blase
threshold $\mu = m_{\Delta,\mathrm{pole}}(0) / 2$ implies a vanishing
curvature mass and the onset of diquark condensation provided that no
first-order transition occurs at lower chemical potentials below the
threshold.

A similar conclusion also holds for arbitrary mesonic pole and curvature masses
carrying vanishing baryon number. However, both masses remain constant
throughout the Silver--Blaze region below the threshold, for example
\begin{equation}
  m^2_{\sigma/\pi,\text{curv}}(\mu) = 
  m^2_{\sigma/\pi,\text{curv}}(0) \; ,
\end{equation}
and
\begin{equation}
  m^2_{\sigma/\pi,\text{pole}}(\mu) = m^2_{\sigma/\pi,\text{pole}}(0) \; .
\end{equation}

\subsection{Diquark two-point functions in RPA}
\label{sec:two-point-functions}

We first analyze the two-point functions of the
  (anti\mbox{-})diquarks at finite real-time frequencies within the
  random-phase approximation (\gls{RPA}). In this framework,
  collective excitations arise from fermionic particle-hole
  scattering, while fluctuations of the composite bosonic fields are
  neglected.  In the context of the \gls{FRG}, the \gls{RPA} two-point
  functions are obtained by computing the two-point functions within
  \gls{LPA} (or \gls{mLPA}) and neglecting all bosonic
  contributions. In this sense, the construction is analogous to
  deriving the \gls{MFA} from the \gls{LPA} flow equation for the
  effective potential, as discussed in \cref{sec:recov-mean-field}.
The \gls{RPA} permits an analytical solution and guarantees a
consistent regularization of the two-point functions with the
underlying mean-field potential. We then improve upon this
approximation by computing the diquark two-point function in
\gls{mLPA}, thereby including mesonic fluctuations.

The flow of the two-point correlation functions for the diquark field
in \gls{RPA} is given by
\begin{equation}
  \label{eq:two_point_flow}
  \partial_t \Gamma^{(2)}_{\Delta_\da^*\Delta_\da,k}(p)
  = \frac{1}{2} \tilde{\partial}_t \Pi_{\Delta_\da^*\Delta_\da,k}(p) \; ,
\end{equation}
where no summation over the index $\da$ is implied. The polarization
loop reads
\begin{align}
  \label{eq:threshold_definition}
    & \Pi_{\Delta_\da^*\Delta_\da,k}(p) \nonumber \\
    & \quad = \int_l \Tr \left[
      \Gamma^{(3)}_{\Delta_\da^*\Psi^{\tp}\Psi} G_{q,k}(l)
      \Gamma^{(3)}_{\Delta_\da\Psi^{\tp}\Psi} G_{q,k}(p+l)
    \right] \; ,
\end{align}
where $\Psi = (q, \bar{q})$.  No four-point vertex $\Gamma^{(4)}$
appears, since bosonic fluctuations are neglected within the \gls{RPA}
(cf. \cref{fig:two_point_flow_lpa}). 
The trace runs over bispinor space as well as color,
flavor, and Dirac indices. The required three-point vertices are
\begin{align}
  \label{eq:23}
  \Gamma^{(3)}_{\Delta_\da\Psi^\tp\Psi} & =
  \begin{pmatrix}
    g_\Delta C\gamma_5 \tau_2 i \epsilon_\da & 0 \\
    0 & 0
  \end{pmatrix} \; , \\[2ex]
  \Gamma^{(3)}_{\Delta_\da^*\Psi^\tp\Psi} & =
  \begin{pmatrix}
    0 & 0 \\
    0 & - g_\Delta C \gamma_5 \tau_2 i \epsilon_\da
  \end{pmatrix} \; .
\end{align}

The \gls{RG}-time derivative on the right-hand side of
\cref{eq:two_point_flow} is defined as
\begin{equation}
  \label{eq:42}
  \tilde{\partial}_t = \int \!\!dx \left[ -2x r'_q(x)  \right]
  \frac{\delta}{\delta r_q (x)}\ ,
\end{equation}
and acts only on the shape function $r_q(x)$ with $x=\sqvec{p}/k^2$,
cf.~\cref{app:regulator-functions}. Both \gls{RG}-scale
  derivatives $\partial_t$ and $\tilde{\partial}_t$ are equivalent
  since the shape function is the only scale-dependent quantity within
  the \gls{MFA} or the \gls{RPA}.  

The flow \cref{eq:two_point_flow} can be integrated from the \gls{UV} scale
$\Lambda'$ down to $k$, yielding a regularized expression for the
diquark two-point function
\begin{equation}
  \label{eq:two-point-flow-solution}
  \begin{aligned}
    \Gamma^{(2)}_{\Delta_\da^*\Delta_\da,k}(p) &= 
    \Gamma^{(2)}_{\Delta_\da^*\Delta_\da,\Lambda'}(p)        \\
    &\quad  + \frac{1}{2} \Pi_{\Delta_\da^*\Delta_\da,k}(p)
    - \frac{1}{2} \Pi_{\Delta_\da^*\Delta_\da,\Lambda'}(p) \; .
  \end{aligned}
\end{equation}
The initial conditions at the \gls{UV} scale $\Lambda'$ are chosen
consistently in terms of the momentum-independent mean-field potential
$U_{k=\Lambda'}$, given in \cref{eq:6}, 
as 
\begin{align} \label{eq:18}
  \Gamma^{(2)}_{\Delta_\da^*\Delta_\da,\Lambda'}(p) = b_1 \; .
\end{align}
This choice corresponds implicitly to a vanishing
  diquark wave-function renormalization at the scale $k=\Lambda'$
  consistent with a composite field introduced via a
  Hubbard-Stratonovich transformation without a kinetic term at the
  \gls{UV} scale.

\subsection{Analytical continuation}
\label{sec:analyt-cont}

To determine the diquark pole mass, we perform an
analytic continuation to real time. At vanishing temperature,
however, the identification of the pole mass does not strictly
require a full analytic continuation with an $i\epsilon$
prescription. In practice, it is sufficient to evaluate the
two-point function at purely imaginary frequencies,
$p_0 = - i \omega$, to extract the vacuum pole mass, while the
$i\epsilon$ prescription becomes relevant only when accessing
real-time properties beyond decay thresholds. In the present case,
the analytic continuation is nevertheless straightforward and is
employed for convenience.

The analytical continuation of the temporal component of the external
momentum in the finite-temperature flow equation for the two-point
function, $\Gamma^{(2)}_{\Delta_\da^*\Delta_\da} (p_0,\vec{p})$, is
implemented in two steps. After performing the Matsubara frequency sum
in the flow equation, the periodicity of the thermal occupation
numbers along the imaginary axis of the complex energy plane is
employed\footnote{In this work, only fermionic distributions are
  needed.}
\begin{equation}
  \label{eq:analytic_continuation_step1}
  n_F(\epsilon + i p_0) \rightarrow n_F(\epsilon) \; ,
\end{equation}
where the external Euclidean Matsubara frequencies are discretized as
$p_0 = 2\pi nT$ as they originate from the diquark field. With this
prescription, the polarization loop \cref{eq:threshold_definition}
takes with $\epsilon_q = \sqrt{\sqvec{l}_{\text{reg}} + m^2_q}$ the
simple form
\begin{align} \label{eq:25}
  \nonumber
  & \Pi_{\Delta^*_\da\Delta_\da,k}(p_0,0) =
  \\ & \qquad \qquad
  16 N_f g_\Delta^2 \int_{\vec{l}}
  \frac{\epsilon_q^2}{4\epsilon_q^2 - (ip_0)^2}
  \frac{-1 + 2 n_F(\epsilon_q)}{\epsilon_q} \; ,
\end{align}
where we have assumed $\mu=0$.

In the second step, the discrete external Matsubara frequency is
analytically continued to the real energy domain according to
$ip_0 \to \omega + i0^+$, which yields the retarded two-point function
for vanishing spatial external momentum components, $\vec{p}=0$
\begin{equation}
  \label{eq:ana_cont}
  \Gamma^{(2),R}_{\Delta_\da^*\Delta_\da}(\omega)
  = - \lim_{\epsilon\rightarrow0} \Gamma^{(2),E}_{\Delta_\da^*\Delta_\da}
  (p_0 = -i(\omega + i\epsilon),0) \; ,
\end{equation}
where $\omega$ is the continuous real frequency.  In practice, the
limit $\epsilon \rightarrow 0$ is not taken analytically but a small
imaginary part, $\epsilon = 10^{-8} \MeV$, is retained.  We have
explicitly verified that the location of the pole masses and the
qualitative features of the spectral function are insensitive to the
choice of $\epsilon$ for $\epsilon \lesssim 0.1\MeV$, see also
Ref.~\cite{Tripolt:2014wra} .

From the two-point function at finite real-time momenta, the pole mass
is immediately accessible, as mentioned in
\cref{sec:silver-blaze-rule}.  The diquark pole mass
$m_{\Delta,\text{pole}}$ is given by the root of the real part of the
retarded two-point function:
\begin{equation}
  \label{eq:pole_mass_def1}
  \text{Re} \, \Gamma^{(2),R}_{\Delta^*\Delta}(\omega= m_{\Delta,\text{pole}}) = 0 \; .
\end{equation}

\subsection{Diquark correlations beyond RPA}

\begin{figure*}[!htb]
  \centering
    \includegraphics[width=\twofigs]{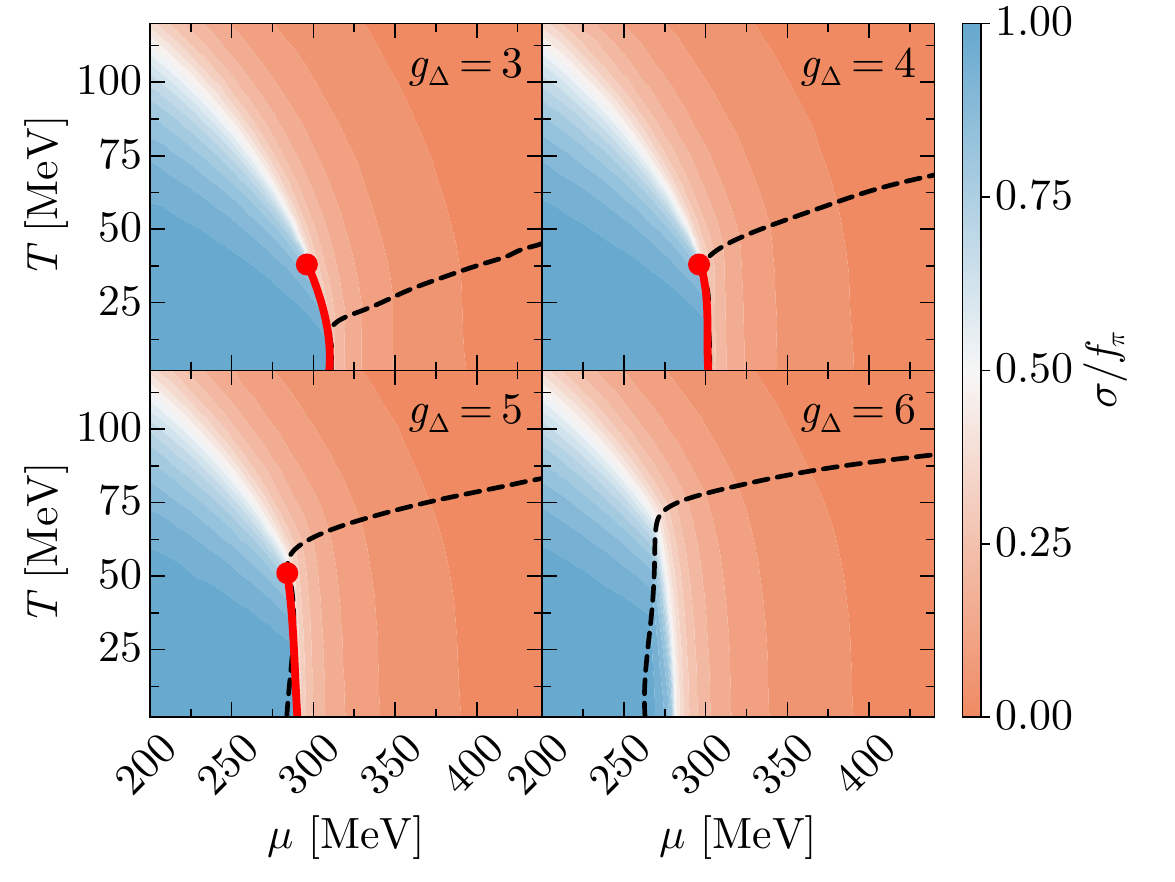}
  \hfill
    \includegraphics[width=\twofigs]{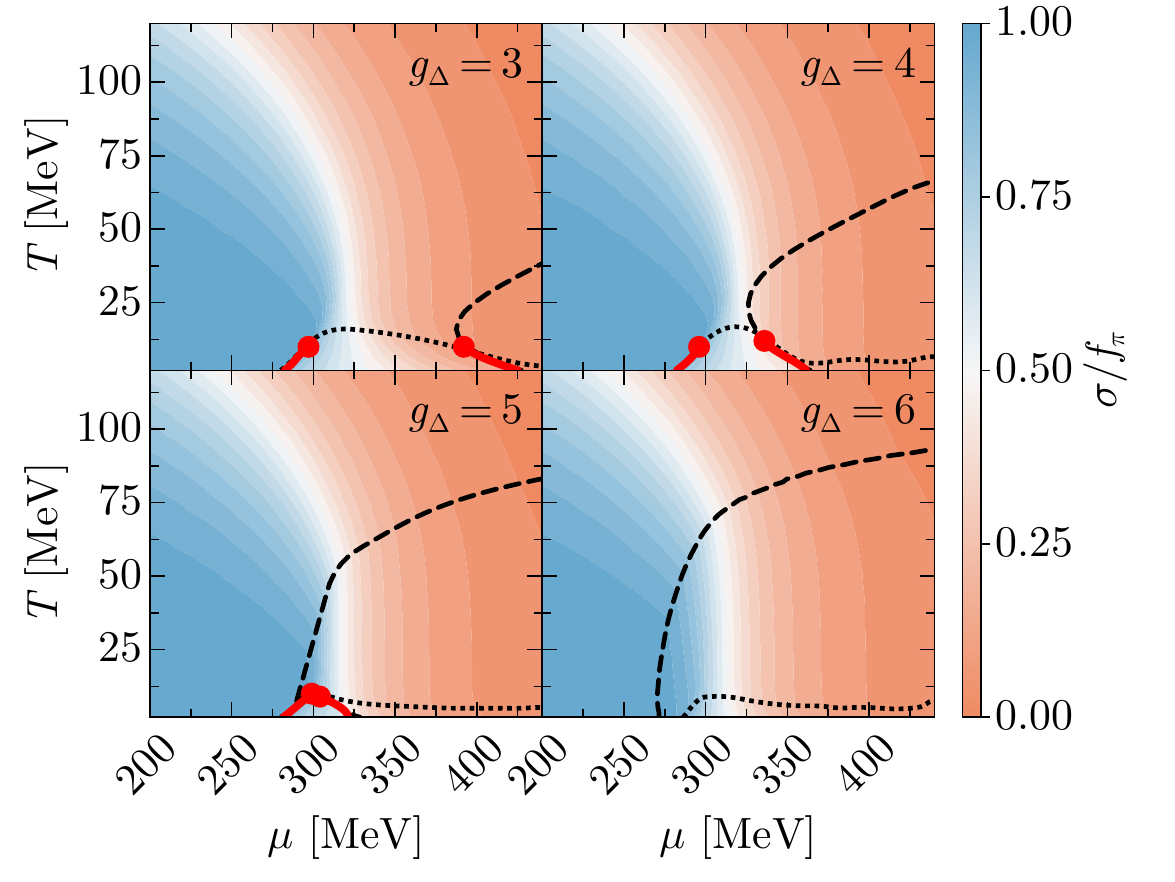}
    \caption{\label{fig:phasediagramQMDmodel} Phase diagrams for the
      quark-meson-diquark model for different diquark couplings
      $g_\Delta$ in \gls{MFA} (left) and \gls{mLPA} (right). The color
      coding represents the strength of the chiral condensate. Red
      solid lines indicate first-order chiral phase transitions
      terminating in a critical endpoint (red dots). In the right
      panel, dotted lines mark the boundary below which the entropy
      density becomes negative.  The black dashed line denotes the
      onset of the diquark condensation. }
\end{figure*}

The diquark two-point function can also be computed in \gls{LPA}.  In
the absence of a diquark condensate, the flow equation of the
two-point function simplifies considerably, which motivates our focus
on this case.

In the \gls{mLPA}, the flow of the diquark two-point function at finite
temperature includes bosonic fluctuations and is shown
diagrammatically in \cref{fig:two_point_flow_lpa}. It is given by
\begin{equation}
  \begin{aligned}
    \partial_t \Gamma_{\Delta^*_\da\Delta_\da,k}^{(2)}(p)
    = \frac{1}{2} \tilde\partial_t \Pi_{\Delta^*_\da\Delta_\da,k}(p)
    + \frac{1}{2} \tilde\partial_t \mathcal{T}_{\Delta^*_\da\Delta_\da,k} \; ,
  \end{aligned}
\end{equation}
where $\mathcal{T}_{\Delta^*_\da\Delta_\da,k}$ denotes a sum of
momentum-independent tadpole diagrams of the form
\begin{equation}
  \label{eq:tadpole_diagrams}
  \mathcal{T}_{\Delta^*_\da\Delta_\da,k}
  = \sum_{\phi=\sigma,\vec{\pi}} \Gamma^{(4)}_{\Delta^*_\da\Delta_\da\phi\phi,k}
  \frac{1}{\beta} \sum_n \int_{\vec{p}}
  G_{\phi\phi,k}(\omega_n, \vec{p}) \; ,
\end{equation}
with bosonic propagators $G_{\phi\phi,k} $ for
$\phi = \left\{ \sigma, \vec{\pi} \right\} $
\begin{equation}
  G_{\phi\phi,k}  (\omega_n, \vec{p}) = \frac{1}{\omega_n^2 +
  \sqvec{p}_{\text{reg,b}} + m^2_{\phi,k}} \; .
\end{equation}
The summation in \cref{eq:tadpole_diagrams} runs over three pions and
one sigma meson and the scale-dependent four-point vertices
$\Gamma^{(4)}_{\Delta^*_\da\Delta_\da\phi\phi,k}$ are related to
derivatives of the meson potential
\begin{align}
  & \Gamma^{(4)}_{\Delta^*_\da\Delta_\da\sigma\sigma,k} = U_k^{(1,1)} +
  \sigma^2 U_k^{(1,2)} \; , \\
  & \Gamma^{(4)}_{\Delta^*_\da\Delta_\da\pi\pi,k} = U_k^{(1,1)} \; .
\end{align}
This identification is valid only in the absence of a diquark
condensate. In this case, all three-point vertices of the form
$\Gamma^{(3)}_{\Delta_\da\phi\phi}$ and
$\Gamma^{(3)}_{\Delta^*_\da\phi\phi}$ vanish, such that bosonic
fluctuations contribute to the diquark two-point flow exclusively via
tadpole diagrams.

Performing the $\tilde\partial_t$ derivative, Matsubara summation, and
momentum integration for the Litim shape function $r_b(x)$ (see
\cref{app:regulator-functions}), the tadpole contribution reduces to
\begin{align}
  \label{eq:46}
  \nonumber
  \frac{1}{2} \tilde\partial_t \mathcal{T}_{\Delta^*_\da\Delta_\da,k}
  = & \frac{k^5}{6\pi^2} \sum_{\phi=\sigma,\pi}
  \frac{ \partial m^2_{\phi,k}} {\partial \Delta^2} \\
  & \times \frac{1}{\epsilon_\phi^2} \left[
    n_B'(\epsilon_\phi) - \frac{1 + 2n_B(\epsilon_\phi)}{2\epsilon_\phi}
  \right] \; .
\end{align}
The bosonic dispersion relations are given in \cref{eq:22} and evaluate
to $\epsilon_\phi = \sqrt{k^2 + m^2_\phi}$.

Numerically, an additional flow equation for $\partial_{\Delta^2} U_k$
is solved and used to determine $\partial_{\Delta^2} m^2_{\phi,k}$ for
all $k$. This allows for a direct integration of the flow equation for
the diquark two-point function at external momentum
$p=(-i(\omega + i\epsilon), \vec{0})$, yielding the retarded two-point
function, analogous to the \gls{RPA}.

\section{Numerical Results}
\label{sec:numerical-results}

We begin by analyzing the phase structure of the \gls{QMD} model in
the mean-field approximation, and compare it to results from the
\gls{FRG}. The impact of diquark pairing on the chiral phase structure
is studied by varying the diquark coupling and examining its
connection to the pole and curvature masses. Unless explicitly stated
otherwise, all results are obtained within the
\gls{RG}-consistent framework.

\begin{figure*}[!tb]
  \centering
    \includegraphics[width=\twofigs]{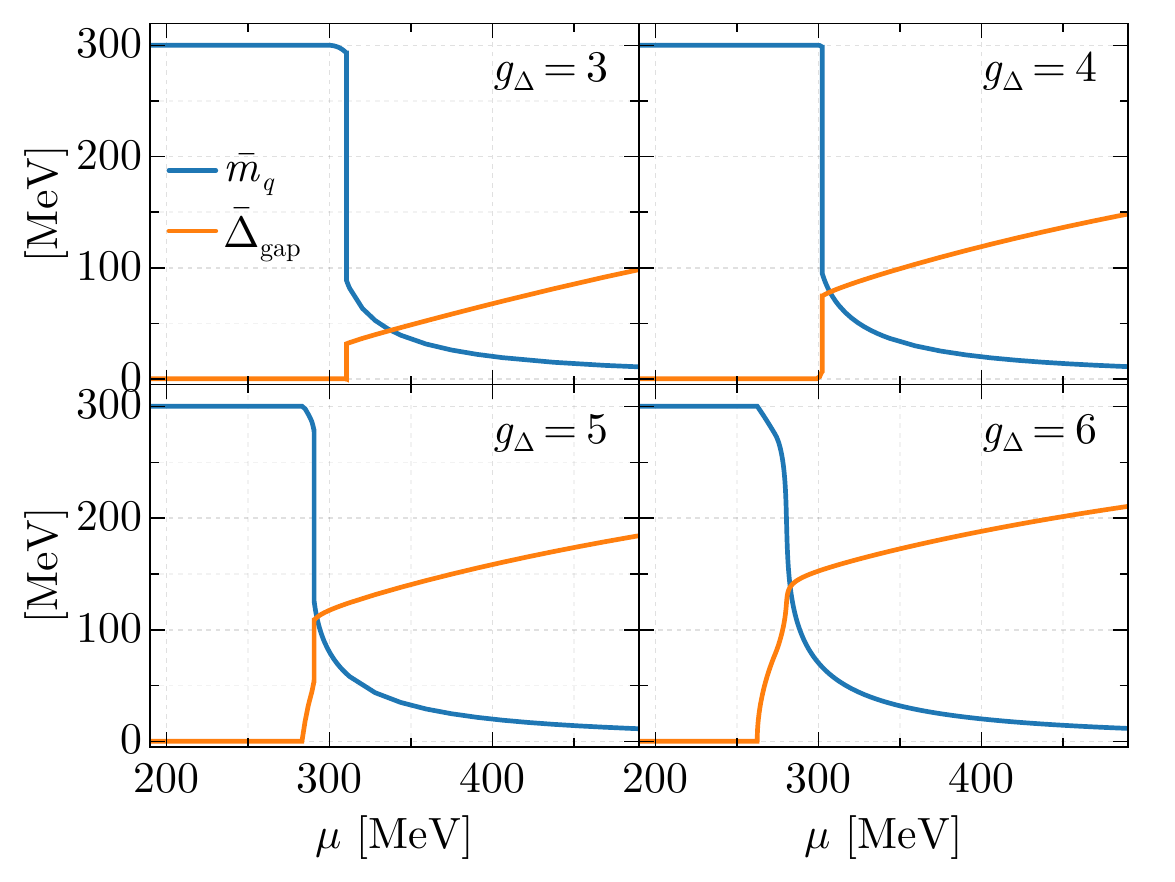}
  \hfill
    \includegraphics[width=\twofigs]{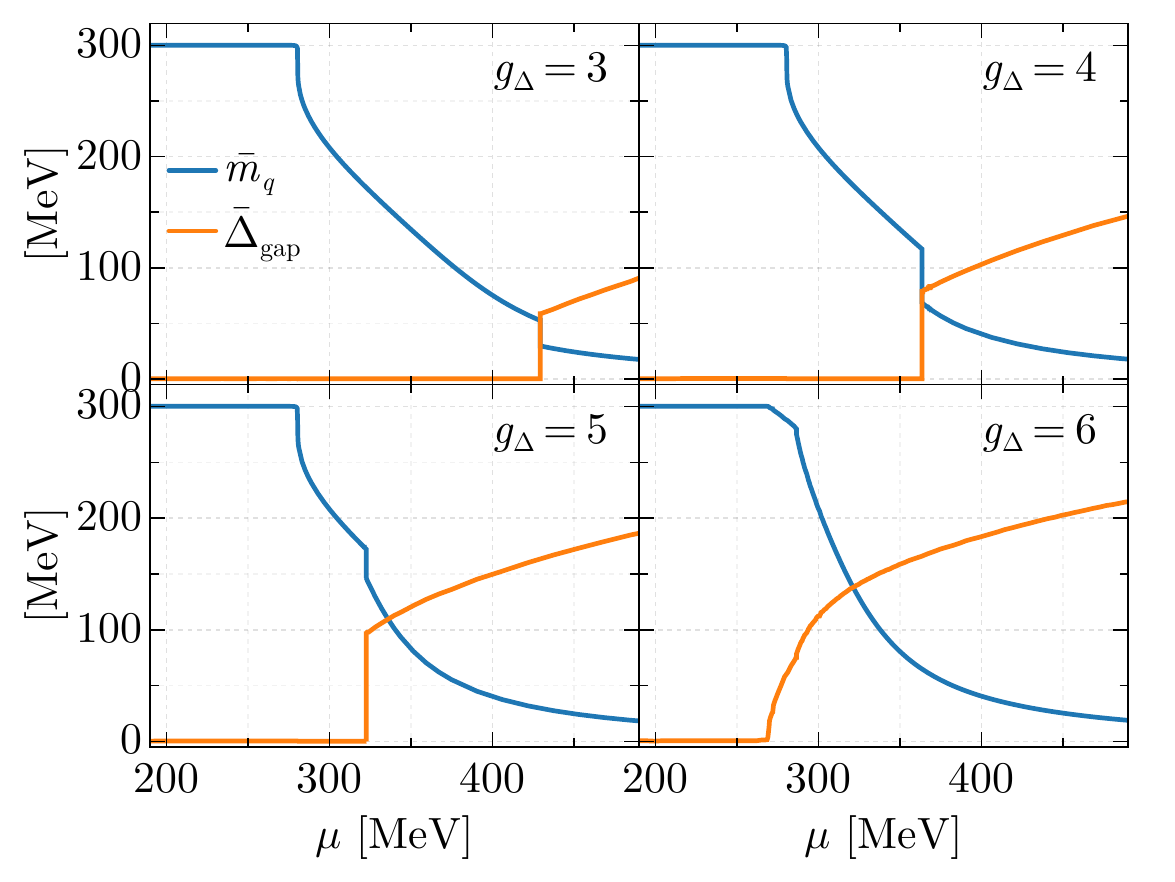}
  \caption{\label{fig:quarkmasses} Quark mass
    $\bar m_{q} = g_\phi\bar\sigma$ and the diquark gap
    $\bar\Delta_{\text{gap}}=g_\Delta \bar\Delta$ for various diquark
    couplings $g_{\Delta}$ as functions of the quark chemical
  potential in \gls{MFA} (left) and \gls{mLPA} (right).}
\end{figure*}

\begin{figure}[!tb]
  \centering
  \includegraphics[width=\columnwidth]{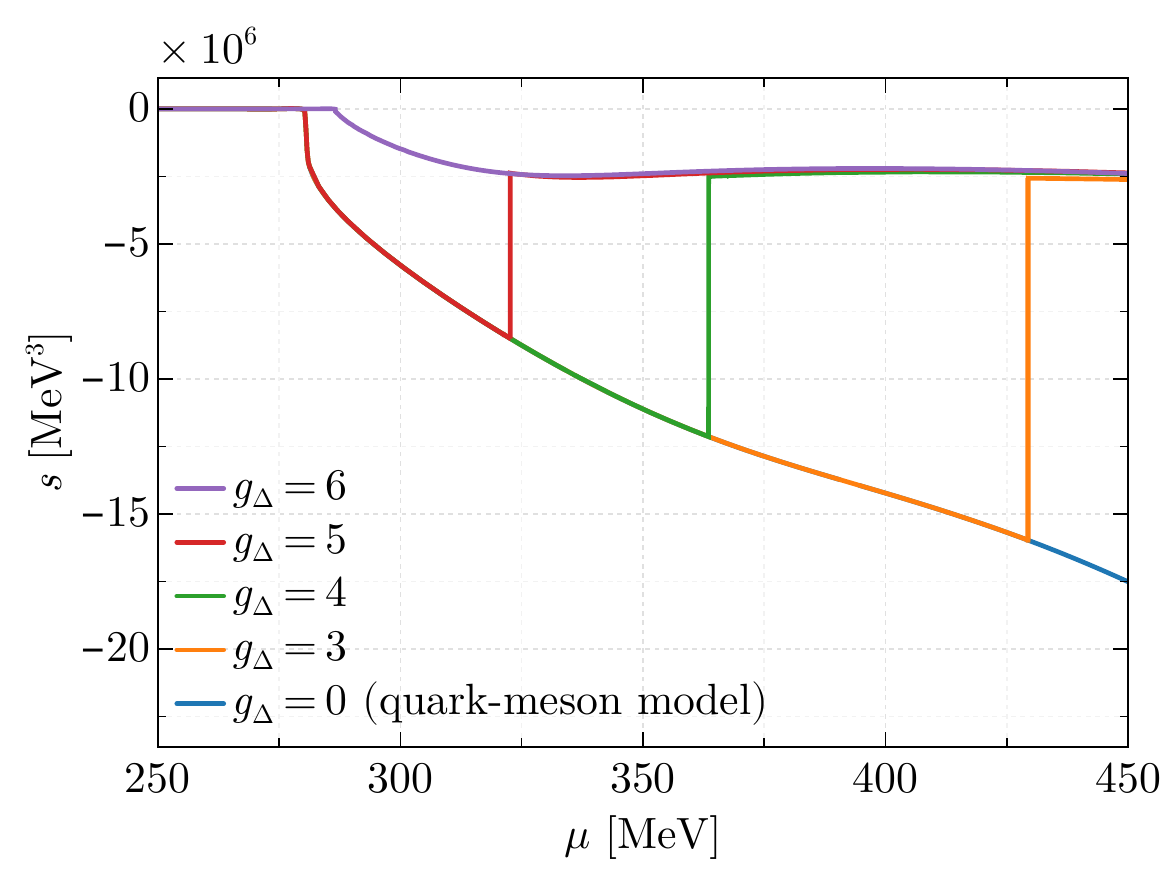}
  \caption{\label{fig:entropydensity} Entropy density $s$ as a
    function of the chemical potential $\mu$ for different diquark
    couplings $g_{\Delta}$ in the \gls{mLPA}, compared with a pure
  quark-meson \gls{FRG} calculation.}
\end{figure}

\subsection{Phase structure}
\label{sec:phase-structure}

The phase structure of the \gls{QMD} model in the $T$-$\mu$ plane,
obtained in \gls{MFA} (left) and in \gls{mLPA} (right) for diquark
couplings $g_\Delta = 3-6$ is shown in \cref{fig:phasediagramQMDmodel}.
The chiral phases are indicated in blue (chirally broken) and orange
(chirally restored) colors. The chiral crossover appears as a white
band, while solid red lines denote the first-order
(chiral) transition terminating in a critical
endpoint (red dot). The onset of the second-order diquark condensation
is indicated by dashed black lines. Additionally, dotted lines in the
right \gls{FRG} results denote the boundary below which the entropy
density becomes negative.

In both \gls{MFA} and \gls{mLPA} results, the overall topology of the
phase diagram remains robust, showing three characteristic regions:
(i) a chirally broken and non-superconducting phase at low $T$ and
$\mu$, (ii) a chirally restored and superconducting phase at low $T$
and high $\mu$, and (iii) a non-superconducting phase at high
$T$. Quantitatively, the \gls{mLPA} shifts the chiral transition lines
to lower chemical potentials, exhibiting a characteristic back-bending
at low temperatures, which reflects the influence of mesonic and
diquark fluctuations. By contrast, in \gls{mLPA} the onset of diquark
condensation is shifted to higher chemical potentials, most
prominently at smaller $g_\Delta$, and is accompanied by a forward
bending of the condensation line. In addition, a
  first-order transition terminating at a critical endpoint appears at
  lower diquark couplings and disappears for larger couplings.

The \gls{MFA} phase diagrams (\cref{fig:phasediagramQMDmodel}, left
panels) show that increasing the diquark coupling raises the critical
temperature $T_c$ at fixed $\mu$, at which diquark pairing is lost
(dashed lines).  This is in line with the \gls{BCS}
relation~\cite{Pisarski:1999tv},
\begin{equation}
  \label{eq:bcs_relation}
  T_c = \frac{e^{\gamma}}{\pi}\bar\Delta_{\text{gap}}(T=0) \; ,
\end{equation}
which links the critical temperature to the zero-temperature diquark
gap, see \cref{fig:quarkmasses}.
Despite deviations from the \gls{BCS} relation at large chemical potentials
in the present \gls{RGC} scheme (see \ccite{Gholami:2025afm}), the critical
temperature remains an increasing function of the zero-temperature
diquark gap.
Thus, \cref{eq:bcs_relation} implies a corresponding rise in $T_c$
with increasing diquark couplings.

This trend persists in the \gls{mLPA} when mesonic fluctuations are
included, but only at higher chemical potentials. Close to the
first-order chiral transition, however, mesonic fluctuations
significantly modify the phase structure in a manner absent in
\gls{MFA}. Interestingly, for larger diquark couplings
($g_{\Delta}= 6 $ in our case) the first-order transition and its
associated endpoint disappear in both \gls{MFA} and \gls{mLPA} calculations.

\subsubsection{Back-bending of the chiral transition line in mLPA}
\label{sec:back-bend-behav}

\begin{figure*}[t]
  \centering
    \includegraphics[width=\twofigs]{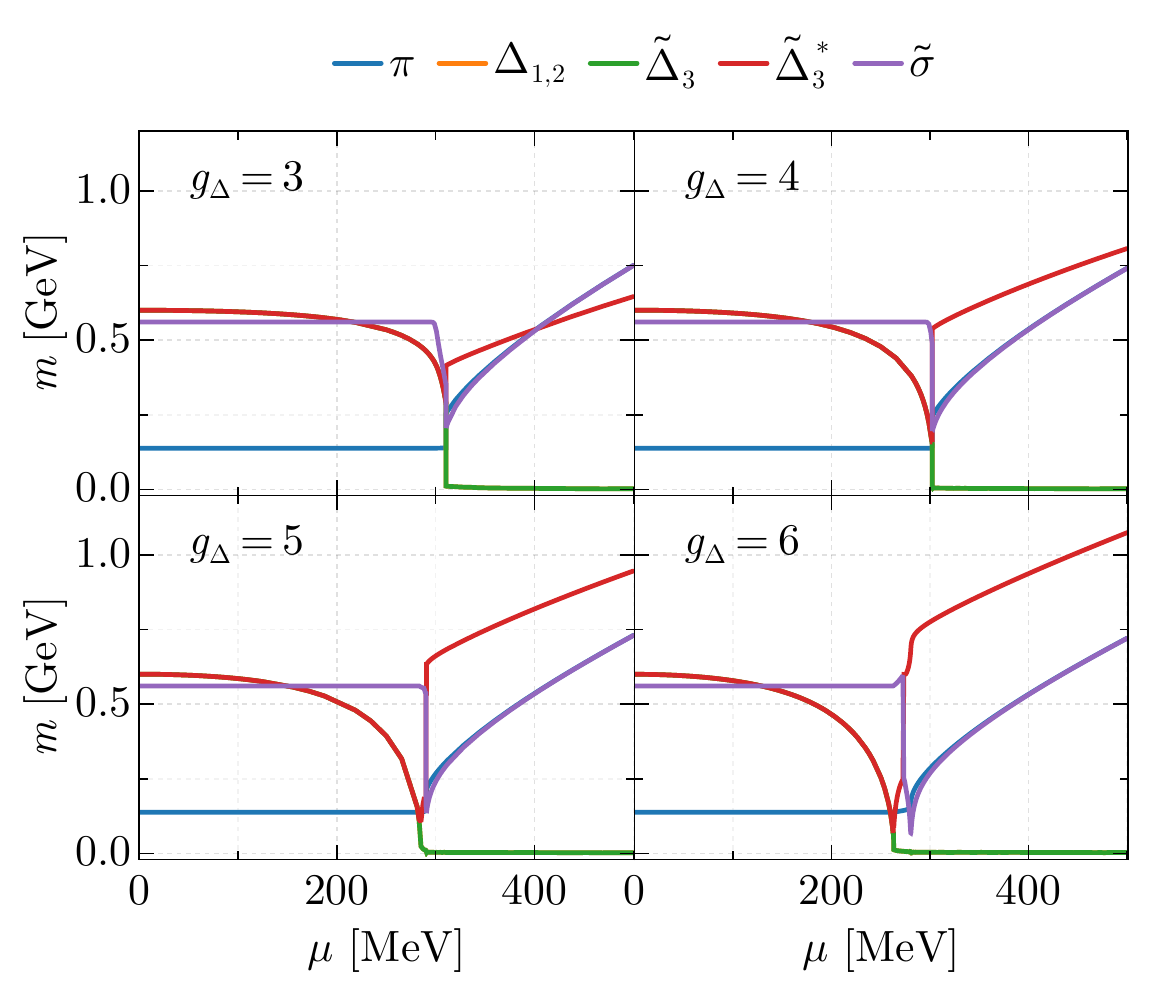}
  \hfill
    \includegraphics[width=\twofigs]{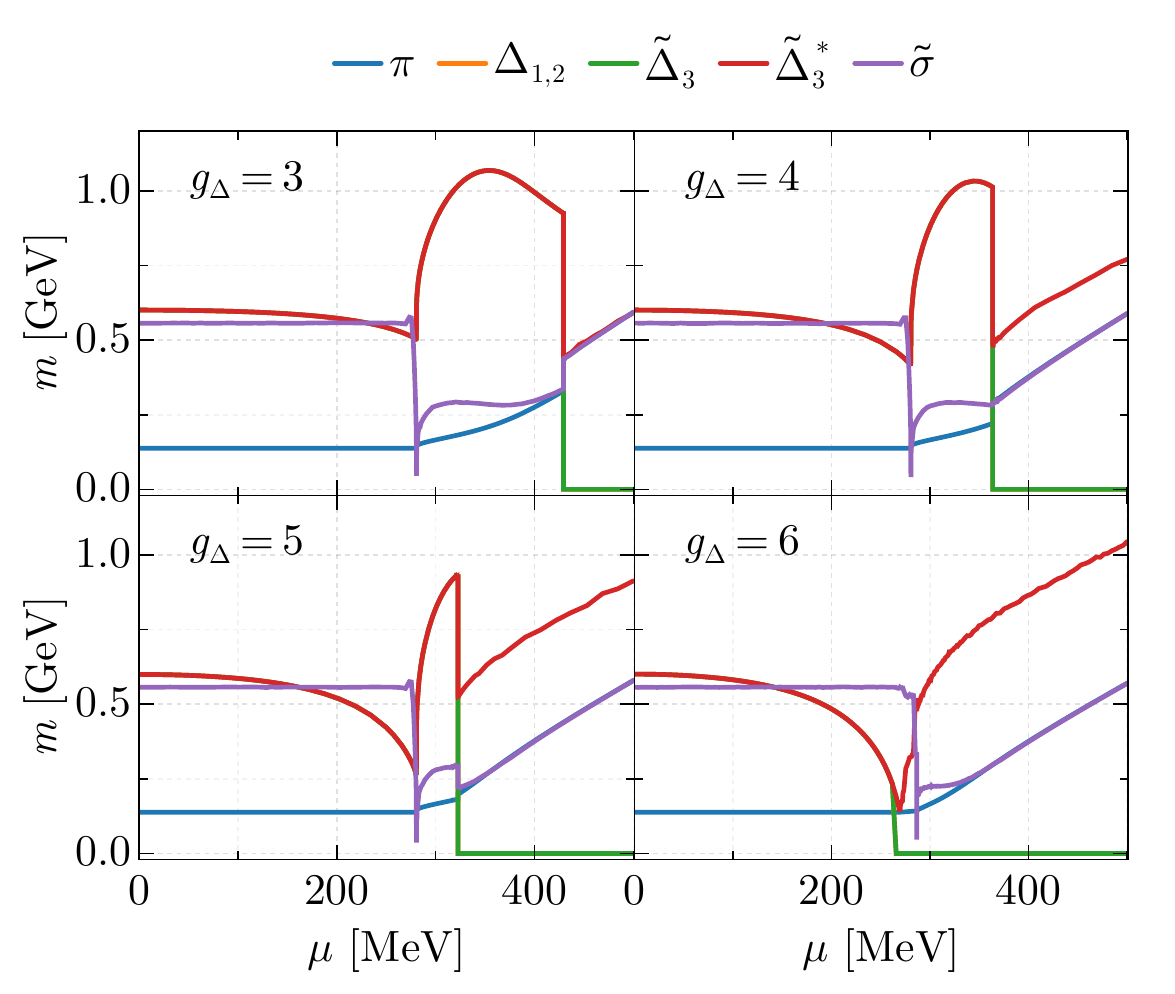}
    \caption{Curvature diquark and meson masses as a function of the
      quark chemical potential $\mu$ for different diquark couplings
      $g_\Delta$ in \gls{MFA} (left) and \gls{mLPA} (right). The
        degeneracy of $\Delta_{1,2}$ and $\tilde\Delta_3$ leads to
        overlapping, indistinguishable curves.
      }
  \label{fig:curvaturemasses}
\end{figure*}

The characteristic back-bending of the chiral transition line at decreasing $T$
(\cref{fig:phasediagramQMDmodel}, right panels) has also been observed in other
low-energy effective model studies beyond \gls{MFA} \cite{Aoki:2015mqa,
Weyrich:2015hha, Tripolt:2017zgc, Zhang:2017icm, Otto:2022jzl, Stoll:2026ezu}, as
well as in mean-field studies of effective Polyakov-loop theories on the lattice
\cite{Konrad:2025ndq}. Due to the Clausius-Clapeyron relation, this back-bending
is related to the appearance of a negative entropy density in this region of the
phase diagram. In \cite{Otto:2022jzl}, the regulator-scheme dependence of the
back-bending was investigated in \gls{LPA}, revealing a sensitivity to the
choice of the shape function. In particular, no negative entropy density was
found for momentum-independent Callan-Symanzik-type regulators, in contrast to
optimized flat regulators.  Since the back-bending is induced at scales around
the Fermi surface and appears as a discontinuity in the fermion flow
\cite{Otto:2022jzl}, it has been conjectured that a Cooper instability of the
Fermi surface driven by attractive pion and sigma exchange may cause the
negative entropy \cite{Tripolt:2017zgc}. This conjecture can now be tested
directly by examining whether the negative-entropy region exhibits
color-superconducting gaps and by assessing the role of diquark condensation in
the back-bending.

As can be seen in \cref{fig:phasediagramQMDmodel}, the back-bending
also affects the diquark onset, which turns into a first-order
transition at low temperature and does not intersect the $\mu$-axis
perpendicularly.  The dotted lines in the figure mark the onset below which
the entropy density becomes negative.  Increasing the diquark coupling
reduces the region of negative entropy, which becomes significantly
suppressed for $g_{\Delta} = 6$. Nevertheless, a finite domain of
negative entropy persists even at large chemical potentials $\mu$.

\cref{fig:entropydensity} shows the entropy density $s$, evaluated in
\gls{mLPA} for different diquark couplings along the chemical
potential axis for (almost) vanishing temperature. Below the chiral
phase transition in the broken phase ($\mu \lesssim 279$ MeV), $s$
vanishes, consistent with the transition location in the corresponding
phase diagram \cref{fig:phasediagramQMDmodel} (right panels). In the
chirally restored phase, the entropy density becomes negative, as
previously observed in a pure quark-meson model analysis
\cite{Tripolt:2017zgc} (lower lines with $g_{\Delta} = 0$ in
\cref{fig:entropydensity}).

\begin{figure*}
  \centering
  \includegraphics[width=\columnwidth]{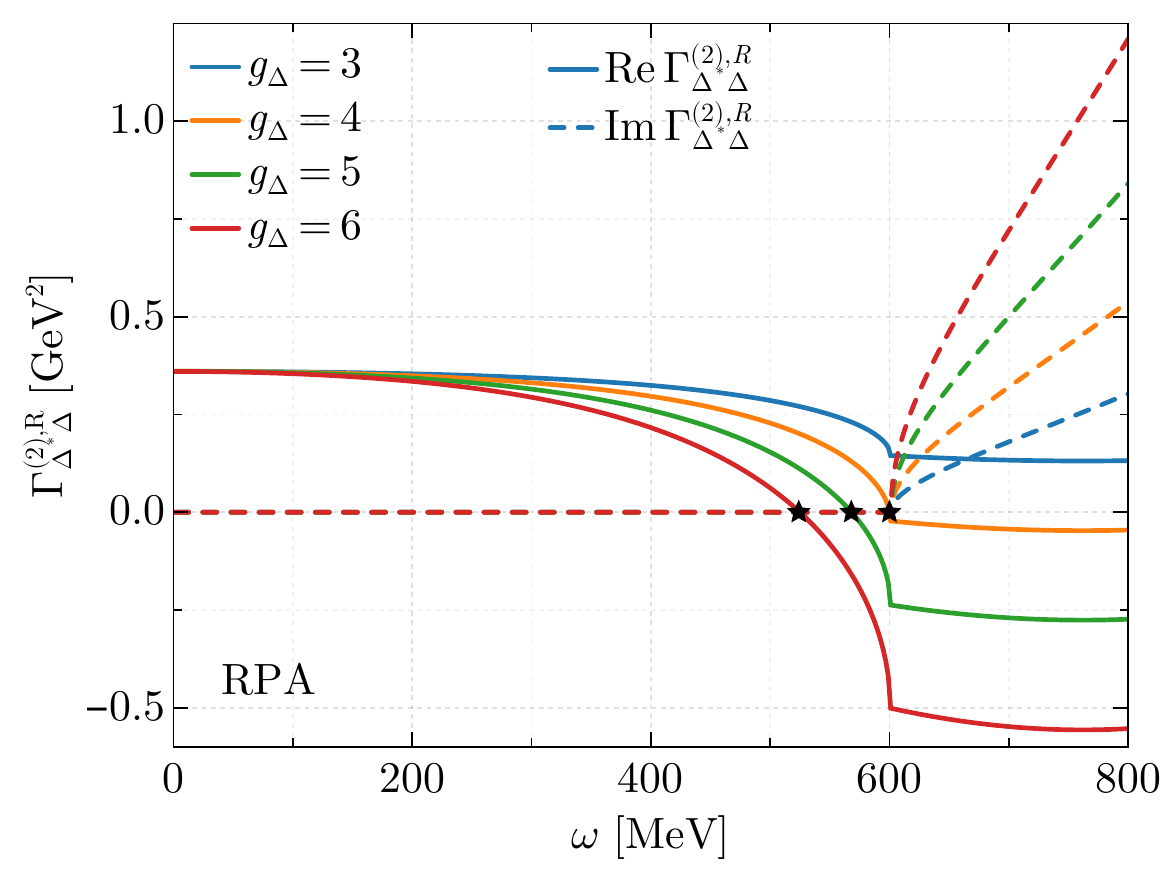}
  \hfill
  \includegraphics[width=\columnwidth]{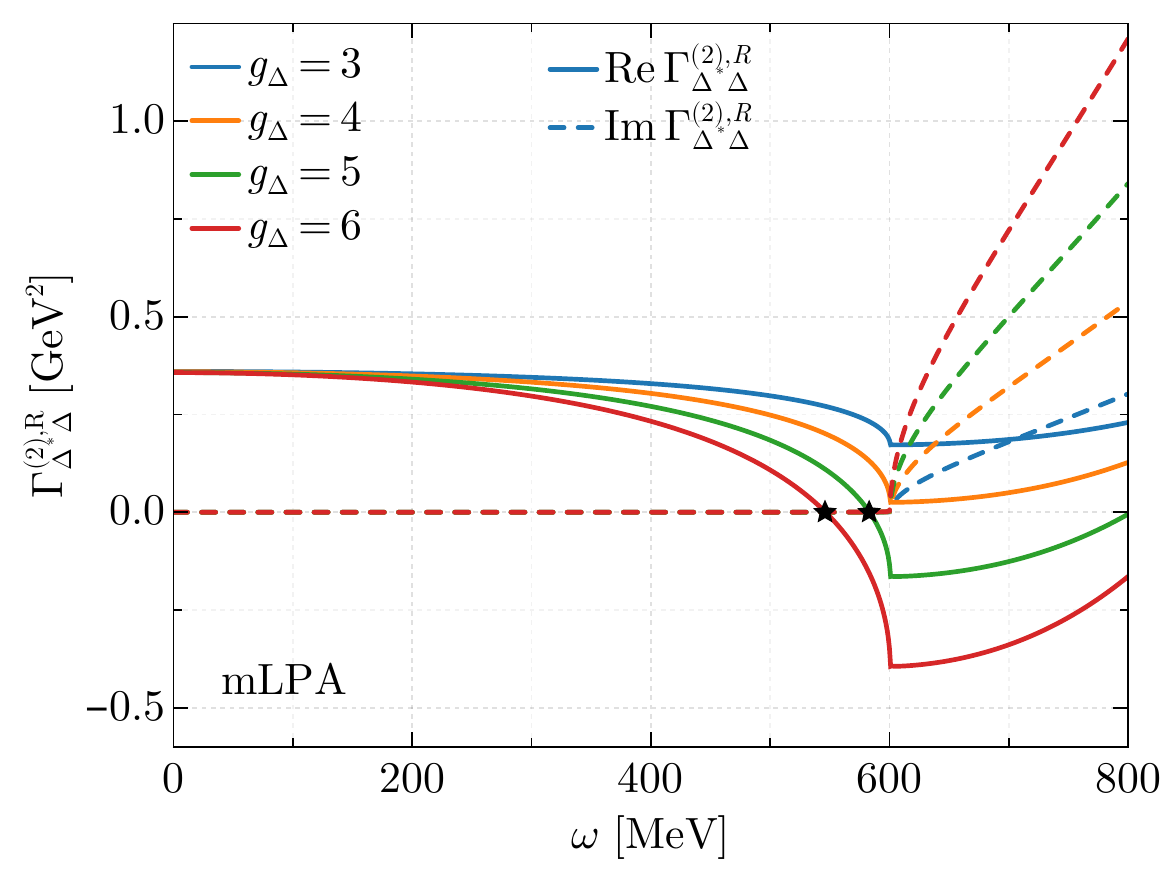}
  \caption{Real (solid lines) and imaginary (dashed lines) parts of
    the vacuum diquark two-point function as a function of the
    real-time external frequency $\omega$ for various diquark
    couplings $g_\Delta$ in \gls{RPA} (left) and \gls{mLPA} (right).
    The black stars mark the locations of the diquark pole masses.
  }
  \label{fig:diquark_two_point_lpa}
\end{figure*}

Once diquark condensation sets in, the entropy density is
significantly reduced by nearly an order of magnitude and remains
approximately constant throughout the \gls{2SC} phase. As the diquark
coupling increases, the onset of condensation shifts to lower chemical
potentials, causing the discontinuities in the entropy associated with
the first-order transition to decrease. For $g_\Delta=6$, these
discontinuities disappear altogether.

We conclude that the inclusion of the diquark condensate at the
mean-field level mitigates the back-bending of the chiral transition
line, leading to a reduced negative entropy density. In particular for
$g_\Delta=6$, this is consistent with
\ccites{Otto:2022jzl,Ihssen:2023nqd}, where the back-bending behavior
diminishes once the chiral transition changes from first to second
order. Furthermore, the contributions of gapped fermions to
thermodynamic quantities with explicit temperature dependence, such as
the entropy density, are exponentially suppressed
\cite{Shovkovy:2002kv}. This suppression may account for the observed
improvements, as the back-bending originates from the feedback of
bosonic fluctuations on the fermionic flow \cite{Otto:2022jzl}.

An open question is whether this trend persists once diquark
fluctuations are included. Clarifying this would help to identify the
origin of the observed improvements. At high densities, the
thermodynamics is expected to be dominated by the diquark sector, as
illustrated in a \gls{MFA} setting in \ccite{Gholami:2025afm}. Since
the diquark channel is handled at the mean-field level, no negative
entropy is expected to arise, leaving it unclear whether the
improvement stems from the diquark channel itself or from its
mean-field treatment. For comparison, we also examined the impact of
the $\omega$-meson on the back-bending and found no significant
modification, consistent with \ccite{Osman:2024xkm}.

\subsubsection{Diquark onset and chiral transition}
\label{sec:diquark_onset}

In \gls{MFA}, the onset of diquark condensation coincides with the
chiral transition at low temperature for $g_{\Delta} =3,4$. For larger
couplings ($g_{\Delta} =5,6$), the diquark onset slightly precedes the
chiral transition, which changes from first order to a sharp crossover
at $g_{\Delta}=6$. This behavior can be seen in \cref{fig:quarkmasses},
where the quark mass and diquark gap are shown as functions of the
chemical potential at (almost) vanishing temperatures.  A similar
pattern is observed in \gls{mLPA}, at least for sufficiently large
diquark couplings. Whenever the diquark onset precedes chiral
restoration, it occurs as a second-order transition at
$\mu < m_q = 300$ MeV for our parameters.  This is consistent with the
Silver-Blaze property, which implies a diquark pole mass smaller than
twice the quark mass, $m_{\Delta,\text{pole}} < 600 \MeV$, as
confirmed below.

\subsection{Pole and curvature masses}
\label{sec:pole_and_curv_mass}

In \cref{fig:curvaturemasses}, we show the curvature masses of mesons
and diquarks as functions of the chemical potential in \gls{MFA} and
\gls{mLPA}.  At low $\mu$, chiral symmetry breaking yields a heavy
sigma meson and three light pions, while the three complex diquark
fields remain degenerate in the absence of a diquark condensate. As
$\mu$ increases within the Silver-Blaze region, $\mu < \mu_c$, the
sigma and pion curvature masses remain constant, whereas the diquark
curvature mass decreases. Both features are fully consistent with the
Silver-Blaze property, see \cref{sec:silver-blaze-rule}.

At larger $\mu$, the system enters the chirally restored and
color-superconducting phase, where the $U(1)_B$ symmetry is broken.
This induces a mixing between the blue diquarks $\Delta_3$,
$\Delta_3^{*}$ and the sigma meson, requiring a diagonalization of the
mass matrix to obtain the in-medium collective modes, denoted by
$\tilde\Delta_3$, $\tilde\Delta_3^*$ and $\tilde\sigma$. As expected
from chiral symmetry restoration, the pion mass becomes degenerate
with one of the collective modes, $\tilde\sigma$.  According to the
symmetry-breaking pattern $SU(3)_c \to SU(2)_c$, we find five
Goldstone modes ($\Delta_1$, $\Delta_1^*$, $\Delta_2$, $\Delta_2^*$,
$\tilde\Delta_3$) and one massive mode ($\tilde\Delta_3^*$), both in
\gls{MFA} and \gls{mLPA}.  The real-time properties of these
excitations at finite density will be addressed in a forthcoming
publication \cite{mire2026prep2}.

In \gls{mLPA}, however, the back-bending region leaves distinct
imprints: at intermediate chemical potentials around the chiral
transition, a non-superconducting phase with partial chiral symmetry
restoration emerges, characterized by a decreasing sigma mass, an
increasing pion mass, and rising (degenerate) diquark masses.

As noted in \cref{sec:diquark_onset}, a second-order onset of diquark
condensation at $\mu<m_{q,\text{vac}}=300\MeV$ requires a vacuum
diquark pole mass below $600 \MeV$.  This is not reflected in the
curvature masses shown in \cref{fig:curvaturemasses}, since our
parameter fixing enforces a vacuum diquark curvature mass of
$600 \MeV$ (see \cref{sec:param-fixing}). To resolve this, we extract
the pole mass from the diquark two-point function evaluated at
real-time external frequencies, cf.~\cref{eq:pole_mass_def1}.

In \cref{fig:diquark_two_point_lpa}, we show the vacuum diquark
retarded two-point function $\Gamma^{(2),R}_{\Delta^*\Delta}(\omega)$
for different diquark couplings in \gls{RPA} and \gls{mLPA},
cf.~\cref{sec:two-point-functions}.  At vanishing frequency, the
retarded two-point function defines the diquark curvature mass
\begin{equation}
  \Gamma^{(2),R}_{\Delta^*\Delta}(\omega = 0) 
  = m^2_{\Delta,\text{curv}}
  = \big( 600 \MeV \big)^2 \; , 
\end{equation}
and hence all computations agree at this point as a consequence of the
parameter-fixing procedure, see \cref{sec:param-fixing}.

\begin{figure*}[!htb]
  \centering
    \includegraphics[width=\twofigs]{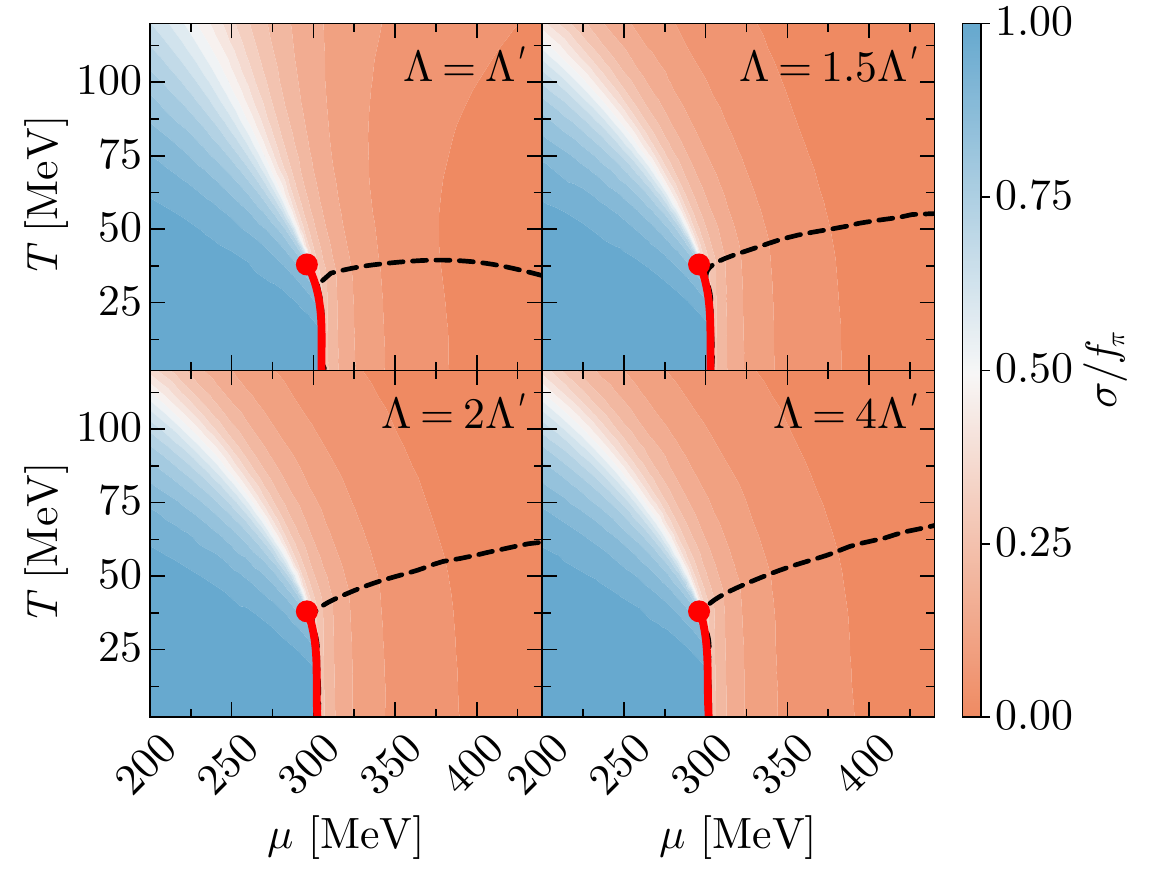}
  \hfill
    \includegraphics[width=\twofigs]{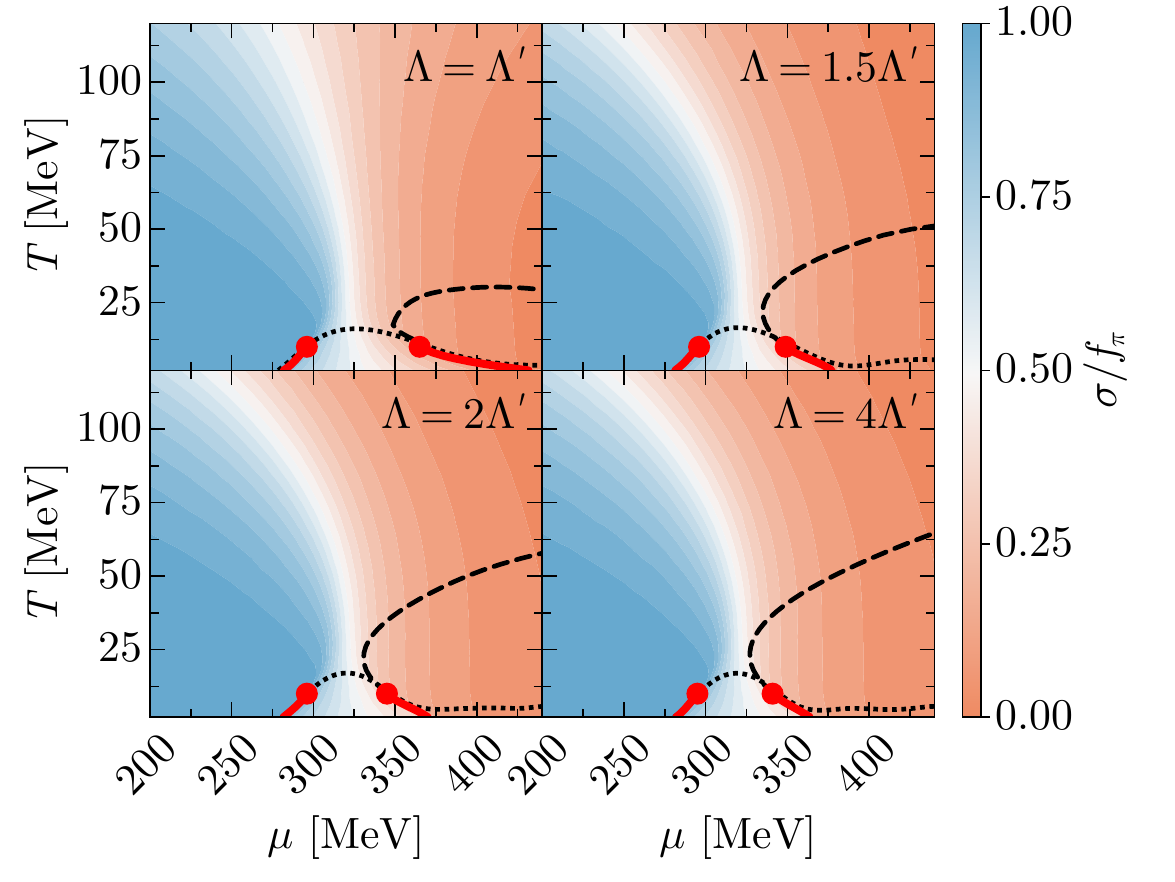}
  \caption{Impact of \gls{RG} consistency on the phase structure of the
    quark-meson-diquark model at fixed diquark coupling
    $g_{\Delta} =4$ in \gls{MFA} (left) and \gls{mLPA} (right).  Color coding
    and line conventions are as in \cref{fig:phasediagramQMDmodel}.
    With increasing \gls{UV} scale $\Lambda$ (at fixed $\Lambda'=600\MeV$),
    cutoff artifacts are reduced in both \gls{MFA} and \gls{mLPA}, while the
    back-bending of the transition lines remains unaffected.
    }
  \label{fig:phase_diagrams_few_rgc}
\end{figure*}

Going to higher external frequencies $\omega$, the real part of the
retarded two-point function starts to decrease. For the largest
diquark couplings ($g_\Delta \geq 4$ in \gls{RPA} and
$g_\Delta \geq 5$ in \gls{mLPA}), the two-point function vanishes
already for $\omega < 2 m_q = 600 \MeV$, corresponding to a finite
diquark pole mass. Increasing the diquark coupling results in a
decreasing pole mass in both approaches and the imaginary part remains
zero up to the threshold at twice the vacuum quark mass,
$\omega= 2 m_q = 600 \MeV$, which marks the onset of the first
available decay channel of the diquark field within the present
approximation.

We find for $g_{\Delta} > 4$ that the diquark pole mass in
\cref{fig:diquark_two_point_lpa} is consistent with twice the onset
chemical potential of diquark condensation. In this regime, diquark
condensation occurs prior to the chiral transition and is realized as
a second-order transition, as shown in
\cref{fig:phasediagramQMDmodel,fig:quarkmasses}. 
This behavior is consistent with the Silver-Blaze property (see
\cref{sec:silver-blaze-rule}), which implies the existence of a diquark pole
mass below $2 m_{q,\text{vac}} = 600 \MeV$ as soon as the chemical potential of
a second-order diquark onset is less than $m_{q,\text{vac}}$.
In particular, we attribute the reduced back-bending behavior observed
in \gls{mLPA} (cf.~\cref{sec:back-bend-behav}) for large diquark
couplings ($g_\Delta=6$) to the finite diquark pole mass found in
\cref{fig:diquark_two_point_lpa} (right).

Since the diquark curvature mass defined in \cref{eq:9} is not an
  \gls{RG}-invariant quantity, a proper comparison between pole and
  curvature mass must also account for the diquark wave-function
  renormalization $Z_\Delta$.  The diquark wave-function
  renormalization and curvature mass are defined through the diquark
  two-point function as \cite{Helmboldt:2014iya}
\begin{equation}
  \label{eq:diquark-wave-function}
  \Gamma_{\Delta^*\Delta}^{(2)}(p) \approx 
  Z_\Delta (m^2_{\Delta,\text{curv}} + p^2) 
  + \mathcal{O}(p^4) \; ,
\end{equation}
which reduces  to \cref{eq:9} for  $Z_\Delta = 1$ and $p=0$.
Using our results for the two-point
function, cf.~\cref{fig:diquark_two_point_lpa}, we extract the corresponding
values of $Z_{\Delta}$ for different couplings, as summarized in
\cref{tab:wavefunction}.
\begin{table}[!h]
\begin{center}
  {
    \renewcommand{\arraystretch}{1.5}
    \begin{tabular}{c @{\hspace{0.25em}} ||
      *{3}{@{\hspace{0.25em}}c|@{\hspace{0.25em}}} c}
      $g_\Delta$ & 3 & 4 & 5 & 6 \\
      \hline
      $Z_\Delta$ & 0.14 & 0.25 & 0.39 & 0.57
    \end{tabular}
    \caption{\label{tab:wavefunction} Diquark wave-function
      renormalization for different diquark couplings. The values
      obtained in \gls{RPA} and \gls{mLPA} differ only by a few percent.
    }
  }
\end{center}
\end{table}

As the diquark coupling decreases, the fermionic contribution to the
diquark wave-function renormalization becomes less pronounced, and
$Z_{\Delta}$ approaches its bare, initial \gls{UV} value
$Z_{\Delta,k=\Lambda}=0$, see \cref{eq:18}. This implies that
including a wave-function renormalization typically enhances the
curvature mass relative to the case without such a correction,
provided that $Z_{\Delta}$ is not treated self-consistently in the
flow.  This suggests that quantitatively relevant
  corrections may arise from a fully self-consistent inclusion of the
  diquark wave-function renormalization, e.g., within an \text{LPA}$'$
  truncation. Such an extension could improve the agreement between
  the diquark pole and curvature masses, analogous to the behavior
  observed for the pion field \cite{Helmboldt:2014iya}.

\subsection{Consequence of RG consistency}

To conclude the discussion of our numerical findings, we assess the
impact of \gls{RGC} on the phase
structure. \cref{fig:phase_diagrams_few_rgc} shows the dependence on
the \gls{UV} cutoff $\Lambda$ at a fixed diquark coupling
$g_\Delta=4$, in analogy to the previous phase diagrams.  In the range
$\Lambda=\Lambda'$ to $\Lambda=1.5\Lambda'$, cutoff artifacts are most
pronounced in both transition lines. The chiral transition at lower
chemical potentials bends towards lower temperatures, while the
diquark condensation line at higher chemical potentials shifts to
higher temperatures as the cutoff is increased.  Increasing the
\gls{UV} cutoff further (to $2\Lambda'$ or $4\Lambda'$ in our case)
leads to no significant changes, highlighting the convergence of the
\gls{RGC} scheme as the \gls{UV} cutoff $\Lambda$ grows well above any
external scale.

This nicely demonstrates that cutoff artifacts, which are present when
external parameters such as $T$ and $\mu$ are near the scale
$\Lambda$, are reduced as $\Lambda$ increases, consistent with
previous studies \ccites{Braun:2018svj, Gholami:2024diy,
  Gholami:2025afm}. Similarly, in the \gls{mLPA} truncation, we
observe analogous improvements compared to the mean-field case, but no
significant modification of the back-bending behavior at low
temperatures. This suggests that the back-bending is not driven by
cutoff artifacts, in agreement with \ccite{Stoll:2026ezu}.

\section{Summary and Conclusions}
\label{sec:summary}

In this work, we present a comprehensive analysis of the phase
structure of the \gls{QMD} model and explore its relation to diquark
two-point functions in the context of the Silver-Blaze property. Our
study is based on a \gls{FRG} approach with two main
objectives. First, it incorporates bosonic fluctuations beyond the
standard \gls{MFA}, which we implemented in the (pseudo)scalar mesonic
channel. Second, by employing the \gls{RG} consistency, it enables a
systematic removal of cutoff artifacts, which is important in
color-superconducting models due to medium-induced
divergences.

In agreement with previous quark-meson model studies, we find that the
inclusion of the (pseudo)scalar mesonic fluctuations
leads to significant modifications of the phase
structure. The characteristic back-bending behavior of the chiral
transition is still present but is strongly affected by the onset 
of diquark condensation.  For
sufficiently large diquark coupling, diquark condensation dominates
the phase structure in both \gls{MFA} and \gls{mLPA}.  
In the latter, this is accompanied by a noticeable reduction of the
back-bending, although it is not completely removed, as reflected, for
example, in regions of negative entropy density.

For large diquark couplings, diquark condensation
sets in below both the vacuum quark mass and the chiral transition
in both \gls{MFA} and \gls{mLPA}.  Since the diquark onset is of second order,
the Silver-Blaze property implies a diquark pole mass below twice the
vacuum quark mass. This is confirmed by an explicit calculation of the
vacuum diquark two-point function in both \gls{MFA} and \gls{mLPA}. We
therefore conclude that the observed reduction of the back-bending at
large couplings is closely linked to the emergence of a finite diquark
pole mass.  Notably, while the diquark curvature mass is fixed to
twice the vacuum quark mass by construction, the pole mass can deviate
significantly from the curvature mass.

The present mean-field treatment of the diquark sector constitutes a
first step towards an \gls{FRG} description of the
\gls{QMD} model. A natural extension is the consistent inclusion of
diquark fluctuations, which is expected to provide further insight
into the origin of the reduced back-bending of the chiral transition
line. At present, it remains unclear whether this reduction is
primarily driven by diquark condensation or reflects an artefact of
the present diquark truncation.

In conclusion, our study highlights the relevance of the vacuum
diquark pole mass in understanding the phase structure of dense quark
matter. Its value relative to the vacuum constituent quark mass can
constrain the location of the diquark onset and the interplay between
chiral symmetry breaking and color superconductivity. Since the
Silver-Blaze property is exact in \gls{QCD}, this connection between
the diquark pole mass and the onset of color superconductivity is
expected to remain robust in more complete \gls{FRG} truncations and
ultimately in full \gls{QCD}. This motivates a first-principles
determination of the scalar diquark vacuum properties, for instance
within lattice \gls{QCD}, more advanced \gls{FRG} truncations, or
\gls{DSE} approaches.

\section*{Acknowledgments}

We thank Michael Buballa, Christian Fischer, Hosein Gholami, Lutz Kiefer, Jan
Pawlowski, Fabian Rennecke, Franz Sattler and Lorenz von Smekal for
discussions. This work has been supported by the Helmholtz Graduate School for
Hadron and Ion Research (HGS-HIRe) for FAIR, the GSI Helmholtzzentrum für
Schwerionenforschung and the Deutsche Forschungsgemeinschaft (DFG, German
Research Foundation) through the Collaborative Research Center TransRegio CRC-TR
211 “Strong-interaction matter under extreme conditions”.

\section*{Data Availability}

The numerical data presented in all figures in this work are openly
available in the ancillary files of the corresponding arXiv
submission.

\appendix

\section{Regulator functions}
\label{app:regulator-functions}

Actual computation of the average action $\Gamma_k$ requires the
explicit specification of the generic regulator $R_k (p,q)$. A
momentum-space representation of the flow allows to evaluate the
functional traces for constant field configurations.  The regulator
should be chosen such that the analytic properties and symmetries of
the underlying theory are respected, as emphasized in studies of
regulator-scheme dependence and \gls{RG} consistency
\cite{Balog:2019rrg,Braun:2020bhy,Otto:2022jzl}. This is achieved by
choosing the same matrix structure as the kinetic term which respects
the Lorentz, Dirac, color and flavor structure of the theory.  The
regulator function is matrix-valued and, depending on the field space,
has diagonal entries for bosonic fields and symplectic ones for
fermions.

We employ spatial three-dimensional flat regulators
\cite{Litim:2001up}.  This choice of regulator does not affect the
frequency component of the 4-momentum and enables the internal
frequency integration, or Matsubara summation at finite temperature,
of the flow equations explicitly. This is crucial for the analytic
continuation of the two-point flow equations from imaginary to real
frequencies, which is necessary for the analysis of the diquark pole
mass. It is furthermore consistent with the Silver-Blaze property
\cite{Braun:2020bhy}. The breaking of Lorentz symmetry induced by
three-dimensional regulator functions is small, well understood, and
does not significantly affect the phase structure or two-point
functions.

For bosonic fields, the regulators read in terms
of the dimensionless momentum ratio $x= \sqvec{p}/k^2$
\begin{equation}
  \label{eq:5}
  R_{k,b} (\sqvec{p})= \sqvec{p}  r_b(x)
\end{equation}
with the dimensionless bosonic shape function
\begin{equation}
  \label{eq:10}
  r_b(x) = (1 / x - 1)\theta(1 - x)\ .
\end{equation}
For fermionic fields,
\begin{equation}
  \label{eq:3d_regulator_fermions}
  R_{k,q} (\vec{p} )= i\slashed{\vec{p}} \; r_q (x)
\end{equation}
the  dimensionless fermionic shape function is given by
\begin{equation}
  \label{eq:litim_shape_function_fermions}
  r_q(x) = (1/\sqrt{x} - 1)\theta (1-x) \ .
\end{equation}
Both shape functions $r_b(x)$ and $r_q(x)$ are related by
\begin{equation}
  \label{eq:37}
  (1 + r_q(x))^2 = (1+r_b(x)) \; ,
\end{equation}
which has been used in the flow \cref{eq:flow_arbitrary_shape}.  In
field space, fermionic regulators always have a symplectic structure.
For example, for our choice of the superfield, \cref{eq:17},
the regulator matrix reads 
\begin{equation}
  \label{eq:36}
  R_k =
  \begin{pmatrix}
    0                             & i\slashed{\vec{p}}^\tp \, r_q (x)
    & 0                    &
    0                             & 0
    \\
    i\slashed{\vec{p}} \, r_q (x) & 0
    & 0                    &
    0                             & 0
    \\
    0                             & 0
    & \sqvec{p} \, r_b (x) &
    0                             & 0
    \\
    0                             & 0
    & 0                    &
    0                             & \sqvec{p} \, r_b (x)
    \\
    0                             & 0
    & 0                    &
    \sqvec{p} \, r_b (x)          & 0
    \\
  \end{pmatrix} \; .
\end{equation}

\section{Quark propagator}
\label{app:propagator_inversion}

In this appendix, we detail the inversion of the quark propagator
$G_{q,k}^{-1}$ defined in \cref{eq:inverse-quark-prop}, including its
nontrivial color structure. Rather than employing the Nambu-\gorkov
formalism to invert the quark propagator, see
e.g.~\ccite{Huang:2001yw}, we adopt a more straightforward approach.
The regularized inverse propagator at finite
  temperature and chemical potential in the field space
$\Psi (p)= (q(p), \bar{q} (p))$ can be expressed in block-matrix form
as
\begin{align}
  \label{eq:54}
  G_{q,k}^{-1} (p)   & =
  \begin{pmatrix}
    \Delta     &  & G_0^{-1, \tp} \\[1ex]
    \bar{G}_0^{-1} &  & \Delta^\dagger
  \end{pmatrix} \; .
\end{align}
The  inverse free quark propagator with $m_q = g_{\phi} \sigma$ reads
\begin{align}
  \label{eq:55}
  G_0^{-1} & =(i \nu_n - \mu )\gamma_4 +  i
  \slashed{\vec{p}}_{\text{reg},q} + m_q \; ,
\end{align}
and the gap matrices are defined
as 
\begin{equation}
  \label{eq:69}
   \begin{aligned}
    \Delta  =& i \Delta_{\text{gap}} C \gamma_5 \tau_2
    \epsilon_3 \; ,\\
    \quad \text{and} \quad
    \Delta^\dagger  = &- i \Delta_{\text{gap}} C
    \gamma_5 \tau_2 \epsilon_3 \; ,
   \end{aligned}
\end{equation}
where $\Delta_{\text{gap}} = g_\Delta\Delta$.  The charged-conjugate
inverse propagator follows from
\begin{equation}
\label{eq:45}
\bar{G}^{-1}_0(p) = - G^{-1}_0(-p)\ .
\end{equation}
For notational simplicity, color indices are suppressed in the
propagator, and we use the shorthand
$\epsilon_3 \equiv (\epsilon_3)_{\da\db} = \epsilon_{3\da\db}$.

This matrix can be inverted by means of the two sets of energy
projection operators $P_{\pm}$ and $\bar{P}_{\pm}$ on positive and
negative energy states of the free Dirac equation for massive quarks
\begin{equation}
  \label{eq:P}
  P_\pm (\vec{p}) = \frac{1}{2}\left(
    \gamma_4 \pm \frac{i
    \slashed{\vec{p}}+m_q}{\epsilon_q (\vec{p})}
  \right)\gamma_4 \; ,
\end{equation}
and correspondingly for massive charge-conjugated antiquarks
\begin{equation}
  \label{eq:Pbar}
  \bar{P}_\pm (\vec{p}) = \frac{1}{2}\left(
    \gamma_4 \pm \frac{i\slashed{\vec{p}}-m_q}{\epsilon_q (\vec{p})}
  \right)\gamma_4 \; ,
\end{equation}
with $\epsilon_q (\vec{p})= \sqrt{\sqvec{p} + m_q^2}$. 

Both sets of
operators are hermitian, i.e., $P^{\dagger}_{\pm} = P_{\pm}$, and they
fulfill the usual projection properties $P_\pm P_\pm = P_\pm$,
$P_\pm P_\mp = 0$, and $P_+ + P_- = 1$ for the quark sector and
similarly for the charge-conjugated antiquark sector $\bar{P}_{\pm}$.
Since $\tr P_{\pm} = \tr \bar{P}_{\pm} =2$, each energy projector
corresponds to two spin orientations of the quark and of the
charge-conjugated antiquark.
The relations $\gamma_4 P_\pm \gamma_4 = \bar{P}_\mp $ and
$\gamma_5 P_\pm \gamma_5 = \bar{P}_\pm$ determine the action of the
Dirac matrices on the energy projectors, with $\gamma_4$ exchanging
and $\gamma_5$ preserving the energy sectors.  This guarantees that
the pairing propagator couples the appropriate particle-hole
components without inducing spurious mixing.

One then finds
\begin{align}
  \label{eq:56}
  G_0^{-1} = \left( i \nu_n + \epsilon_q^- \right) P_+ \gamma_4
  + \left( i \nu_n - \epsilon_q^+ \right) P_- \gamma_4 \; , 
\end{align}
with $\epsilon_q^{\pm} = \epsilon_q \pm \mu$. Note that the inverse propagators,
$G_0^{-1}$ and $\bar{G}_0^{-1}$, depend on the regularized momenta
$\vec{p}_{\text{reg},q}$ rather than on $\vec{p}$.

The full inverse quark propagator can be written as
\begin{equation}
  \label{eq:57}
  \mathcal{G}_{q,k}^{-1} = T G_{q,k}^{-1}
  \quad \text{with} \quad
  T =
  \begin{pmatrix}
    0 & 1 \\ 1 & 0
  \end{pmatrix} \; ,
\end{equation}
where the matrix  $T$ provides a convenient rearrangement of the
basis; the same results can be obtained without introducing $T$.
The propagator can then be inverted
straightforwardly \cite{Wang:2005zb}, yielding
\begin{equation}
  \label{eq:58}
  \mathcal{G}_{q,k} =
  \begin{pmatrix}
    G^+   & F^-                  \\
    F^+ & {G^-}^\tp
  \end{pmatrix} \; ,
\end{equation}
with
\begin{align}
  \label{eq:b18}
  [G^+]^{-1}       & = \bar{G}_0^{-1}
  - \Delta^\dagger  G_0^\tp \Delta \; ,    \\
  \label{eq:b19}
  [G^-]^{-1, \tp} & = G_0^{-1, \tp}
  - \Delta \bar{G}_0 \Delta^\dagger \; ,             \\
  \label{eq:b20}
  F^+                    & = - G_0^\tp \Delta G^+ \; , \\
  \label{eq:b21}
  F^-                    & = - \bar{G}_0 \Delta^\dagger  G^{- \tp} \; .
\end{align}
The evaluation of \crefrange{eq:b18}{eq:b21} relies on the
color-space relation
\begin{equation}
  \label{eq:epsilon3_squared}
  (\epsilon_3)^2 = - P_{rg} = -\diag (1,1,0)\; ,
\end{equation}
which gives rise to the characteristic 2SC pairing structure. The transposed
expressions in Dirac space can be simplified using
$C\gamma_5 (P_{\pm}\gamma_4)^\tp C\gamma_5 = P_{\pm}\gamma_4$ and
$C\gamma_5 (\bar{P}_{\pm}\gamma_4)^\tp C\gamma_5 =
\bar{P}_{\pm}\gamma_4$.

One then finds 
\begin{align}
  \label{eq:59}
  G^+ & = G_\Delta P_{rg} + G_0 P_b \; , \\
  G^- & = \bar{G}_\Delta P_{rg} + \bar{G}_0 P_b \; , \\
  F^+ & = F^{-\dagger} = \Xi P_{rg} \; ,
\end{align}
with the propagator in the blue sector, $P_b = \diag(0,0,1)$, given by
\begin{align}
  \label{eq:40}
  G_0 =&
  \frac{1}{i \nu_n - \epsilon_q^-} P_+ \gamma_4
  + \frac{1}{i \nu_n + \epsilon_q^+} P_- \gamma_4 \; ,
\end{align}
and in the red-green sector by
\begin{equation}
  G_\Delta =
  \frac{i \nu_n + \epsilon_q^-}{(i\nu_n)^2 - E_q^{-2}} P_+ \gamma_4
  + \frac{i \nu_n - \epsilon_q^+}{(i\nu_n)^2 - E_q^{+2}} P_- \gamma_4 \; . \\
\end{equation}
Similar to the inverse propagators in \cref{eq:54}, the corresponding
charge-conjugate propagators $\bar{G}_0$ and $\bar{G}_\Delta$ are
related to $G_0$ and $G_\Delta$ via
\begin{equation}
  \label{eq:35}
  \bar{G}_0(p) = - G_0(-p) \; ,
  \quad \text{and} \quad
  \bar{G}_\Delta(p) = - G_\Delta(-p) \; .
\end{equation}
Finally, the full inverted quark propagator $G_{q,k}$, as introduced
in \cref{eq:nambu_gorkov_propagator}, can be obtained via the basis
reordering transformation
\begin{equation}
  \label{eq:39}
  G_{q,k} = \mathcal{G}_{q,k} T \; .
\end{equation}

\section{Initial UV parameters and numerical details}
\label{app:uv-param-numer}

Here, we summarize the numerical input parameters used throughout this
work. Within the \gls{RGC} setup two ultraviolet cutoffs are
employed. The standard \gls{UV} cutoff is denoted by $\Lambda'$, while the
\gls{RGC} \gls{UV} cutoff is taken as $\Lambda=10\Lambda'=6\GeV$, unless
stated otherwise.

\begin{table}[H]
  \centering
  {  
    \renewcommand{\arraystretch}{1.25}
    \begin{tabular}{c c c c c c c c}
      & $b_1 [\MeV^2]$ & $g_\Delta$ & $a_1 [\MeV^2]$ & $a_2$ &
      $g_\phi$ & $h [\MeV^3]$ & $\Lambda' [\MeV]$ \\ \hline\hline
      \multirow{4}{*}{\textbf{MFA}} & $(755.65)^2$ & 6 & $(270.76)^2$
      & 1.70  & 3.25 & $(120.70)^3$ & 600 \\
      & $(712.16)^2$ & 5 & " & " & " & " & " \\
      & $(673.92)^2$ & 4 & " & " & " & " & " \\
      & $(642.65)^2$ & 3 & " & " & " & " & " \\ \hline
      \multirow{4}{*}{\textbf{LPA}} & $(746.32)^2$ & 6 &
      $-(109.34)^2$  & 4.60  & 3.25 & $(120.70)^3$ & 600 \\
      & $(704.27)^2$ & 5 & " & " & " & " & " \\
      & $(669.40)^2$ & 4 & " & " & " & " & " \\
      & $(640.31)^2$ & 3 & " & " & " & " & " \\
  \end{tabular}}
  \caption{Initial \gls{UV} parameters used in the \gls{FRG} and \gls{MFA} analyses.}
  \label{tab:numerical_parameters}
\end{table}

The initial flow parameters at the \gls{UV} scale $\Lambda' = 600\MeV$ are
listed in \cref{tab:numerical_parameters}. As discussed in
\cref{sec:param-fixing}, the parameters in the (pseudo)scalar
interaction channel, $a_1$, $a_2$, $g_\phi$, and $h$, are fixed from
vacuum phenomenology and are taken from Ref.~\cite{Otto:2019zjy}. The
diquark mass parameter $b_1$ is adjusted for each value of the diquark
coupling $g_\Delta$ such that the diquark curvature mass in the
infrared satisfies $m^2_{\Delta,\text{curv}} = (600 \MeV)^2$.

With the input parameters specified above, the numerical determination
of the chiral and diquark condensates is performed by solving the flow
equation \cref{eq:full_flow}, on a one-dimensional $\sigma$ grid using
the finite-difference upwind scheme described in
\ccite{Ihssen:2023qaq}, for fixed values of $\Delta$.  The minimum in
the $\sigma$ direction,
\begin{equation*}
  \min_{\sigma} \big\{ U_{k=0} (\sigma,\Delta) - h\sigma \big\}
\end{equation*}
is then minimized with respect to $\Delta$.

The numerical calculations presented in this work were performed using
the \texttt{SciML} ecosystem \cite{rackauckas2017differentialequations,
vaibhav_kumar_dixit_2023_7738525,pal2024nonlinearsolve}
 within the \texttt{Julia} programming language
\cite{Julia-2017}, supplemented by selected algorithms from the
\texttt{SUNDIALS} suite \cite{gardner2022sundials,
hindmarsh2005sundials}. All figures are produced using \texttt{Makie.jl}
\cite{DanischKrumbiegel2021} and \texttt{TikZ-Feynman}
\cite{Ellis:2016jkw}.

\bibliography{arxiv_v1}

\end{document}